\documentclass[conference]{IEEEtran}
\IEEEoverridecommandlockouts
\usepackage{subfigure}
\usepackage{cite}
\usepackage{amsmath,amssymb,amsfonts}
\usepackage{algorithmic}
\usepackage{algorithm}
\usepackage{graphicx}
\usepackage{textcomp}
\usepackage{xcolor}
\usepackage{url}
\usepackage{bm}
\usepackage{multirow}
\usepackage{makecell}
\usepackage{booktabs}

\usepackage{caption}

\begin{document}

\title{Large Language Model-Enhanced Multi-hop Parallel Image Semantic Communication}
\author{Bingyan Xie, Jihong Park,~\IEEEmembership{Senior Member,~IEEE,} Rui Mao, Longyu Zhou, \\ Tianhao Liang, Yongpeng Wu,~\IEEEmembership{Senior Member,~IEEE,} Wenjun Zhang,~\IEEEmembership{Fellow,~IEEE}
	
\thanks{An earlier version of this paper was accepted by IEEE ICC 2026. \cite{mhpsc}}
\thanks{(Corresponding author: Yongpeng Wu)}
\thanks{Bingyan Xie, Rui Mao, Yongpeng Wu, and Wenjun Zhang are with the Department of Electronic Engineering, Shanghai Jiao Tong University, Shanghai 200240, China (e-mail:bingyanxie, maorui2002, yongpeng.wu, zhangwenjun@sjtu.edu.cn).}
\thanks{Tianhao Liang is with the School of Information Science and Technology, Harbin Institute of Technology, Shenzhen, China (email: liangth@hit.edu.cn)}
\thanks{Jihong Park and Longyu Zhou are with the ISTD Pillar, Singapore University of Technology of Design, 8 Somapah Rd, Singapore 487372 (e-mail:jihong\_park@sutd.edu.sg, zhoulyfuture@outlook.com)}
}
\maketitle
\begin{abstract}
This paper proposes a large language model-enhanced multi-hop parallel image semantic communication (LLM-MHPSC) framework to mitigate distortion accumulation in multi-hop wireless image transmission. Unlike conventional single-hop semantic communication schemes, LLM-MHPSC deploys an extra residual compensation link at each hop to counteract accumulated distortions. To minimize additional bandwidth overhead, a coarse-to-fine residual compression scheme is designed by integrating a deep learning-based compressor with adaptive arithmetic coding (AAC). Furthermore, a large language model-based residual transmission optimizer (LLM-RTO) is developed to accurately estimate residual distributions and enable channel state and hop-aware rate adjustment, thereby improving residual compression efficiency under varying channel and hop conditions. An adaptive hop selection strategy is also proposed to activate the residual link on demand, striking a balance between transmission performance and computational cost. Experimental results show that LLM-MHPSC outperforms state-of-the-art semantic communication and traditional schemes, realizing robust image transmission with a marginal increase in bandwidth. This framework provides a flexible and effective solution for extending semantic communication to practical multi-hop application scenarios.
\end{abstract}

\begin{IEEEkeywords}
semantic communication, large language model, deep learning, distortion accumulation
\end{IEEEkeywords}

\section{Introduction}
Semantic communication has emerged as a promising 6G paradigm for efficient multimedia transmission by shifting from traditional separated source-channel coding (SSCC) toward deep learning (DL)-enabled communication system designs. By transmitting semantic representations rather than raw data, it reduces communication overhead and supports applications such as Internet of Things, smart cities, and extended reality. For wireless image transmission, JSCC-based methods have been widely studied, with advances in SNR adaptation, CSI-aware encoding, and semantic representation learning \cite{ADJSCC,LCFSC,robust,FLSC,genai,contrast}.

Despite these advances, most semantic communication schemes are designed for single-hop transmission, whereas practical systems such as broadcasting networks, unmanned aerial vehicles, and wireless sensing often rely on multi-hop delivery. Directly extending single-hop JSCC to multi-hop scenarios may cause severe distortion accumulation, since successive wireless transmission processes are inherently lossy. Existing studies have explored several solutions for multi-hop semantic transmission \cite{relay,joint,mhdeepsc}. Relay-assisted schemes introduce intermediate nodes to enhance feature forwarding, with hyperprior entropy recompression improving the reliability of the first-hop representation \cite{relay}. Recursive training further enables semantic codecs to account for distortion propagation across successive hops \cite{mhdeepsc}. In addition, joint compression-transmission-computation optimization considers system-level objectives, including end-to-end latency and reconstruction quality, for more efficient multi-hop delivery \cite{joint}.

Although end-to-end optimization schemes \cite{mhdeepsc,joint} partially alleviate distortion accumulation, reconstruction quality still degrades as the hop count increases. This is because identical codec structures and weights are shared across nodes, making later hops sensitive to distortion inherited from preceding links. Therefore, training-based improvements alone may be insufficient to fully overcome the architectural limitations of single link transmission in long multi-hop chains. A promising direction is to introduce a dual-link structure to complement the semantic transmission path. Existing dual-link schemes combine semantic and bit-level streams or improve controllable fidelity in generative semantic communication \cite{parasc,hsc}. However, they are mainly designed for single-hop image delivery, where the auxiliary stream refines only the final reconstruction. In multi-hop scenarios, repeated decoding and re-encoding can still propagate residual errors across relay nodes. To address this issue, residual-aided transmission link injects compressed residuals at each receiving node, using coarse-to-fine residual compression with a DL-based compressor and adaptive arithmetic coding (AAC) to limit additional overhead \cite{mhpsc}.

Despite the potential of residual-aided dual-link transmission, its residual estimation module remains limited in multi-hop scenarios, restricting the compression efficiency of AAC. This mainly arises from convolutional neural network (CNN)-based distribution modeling, which is less effective in capturing long-range dependencies and contextual correlations across hops. As a result, the residual link cannot fully adapt to multi-hop distortion evolution. Large language models (LLMs) offer a promising solution due to their strong contextual modeling and representation learning capabilities. Recent studies have applied LLMs to channel prediction, multi-user beamforming, and integrated sensing and communication, showing their ability to model complex dependencies across heterogeneous modalities and time scales \cite{llm4cp,llmbo,wiregpt,llmsense,llmlow}. These properties are particularly important for multi-hop semantic communication, where residual statistics are jointly affected by image semantics, channel states, and hop-dependent distortion propagation. Motivated by this, we introduce a fine-tuned LLM into the residual compensation link to improve residual distribution estimation and enable CSI and hop-aware rate adaptation. 

Building on the above motivations, we propose LLM-MHPSC, an LLM-enhanced multi-hop parallel image semantic communication (LLM-MHPSC) framework for robust wireless image transmission. To explicitly suppress distortion accumulation, LLM-MHPSC introduces an extra residual compensation link that injects compressed residual information at each hop. To control the additional communication overhead, a coarse-to-fine residual compression scheme is developed by combining a DL-based compressor with AAC. To further improve residual link efficiency, we design an LLM-enhanced residual transmission optimizer (LLM-RTO). By exploiting the contextual modeling capability of LLMs, LLM-RTO improves residual distribution estimation for AAC and enables CSI and hop-aware rate adjustment. Moreover, an adaptive hop selection strategy is introduced to activate the residual link only when necessary, balancing reconstruction quality and computational cost. Through these designs, LLM-MHPSC provides an efficient and flexible solution for mitigating multi-hop distortion accumulation with affordable bandwidth and computation overhead. The main contributions are summarized as follows.

\begin{figure*}[htbp]
	\centering
	\includegraphics[width=6.2in]{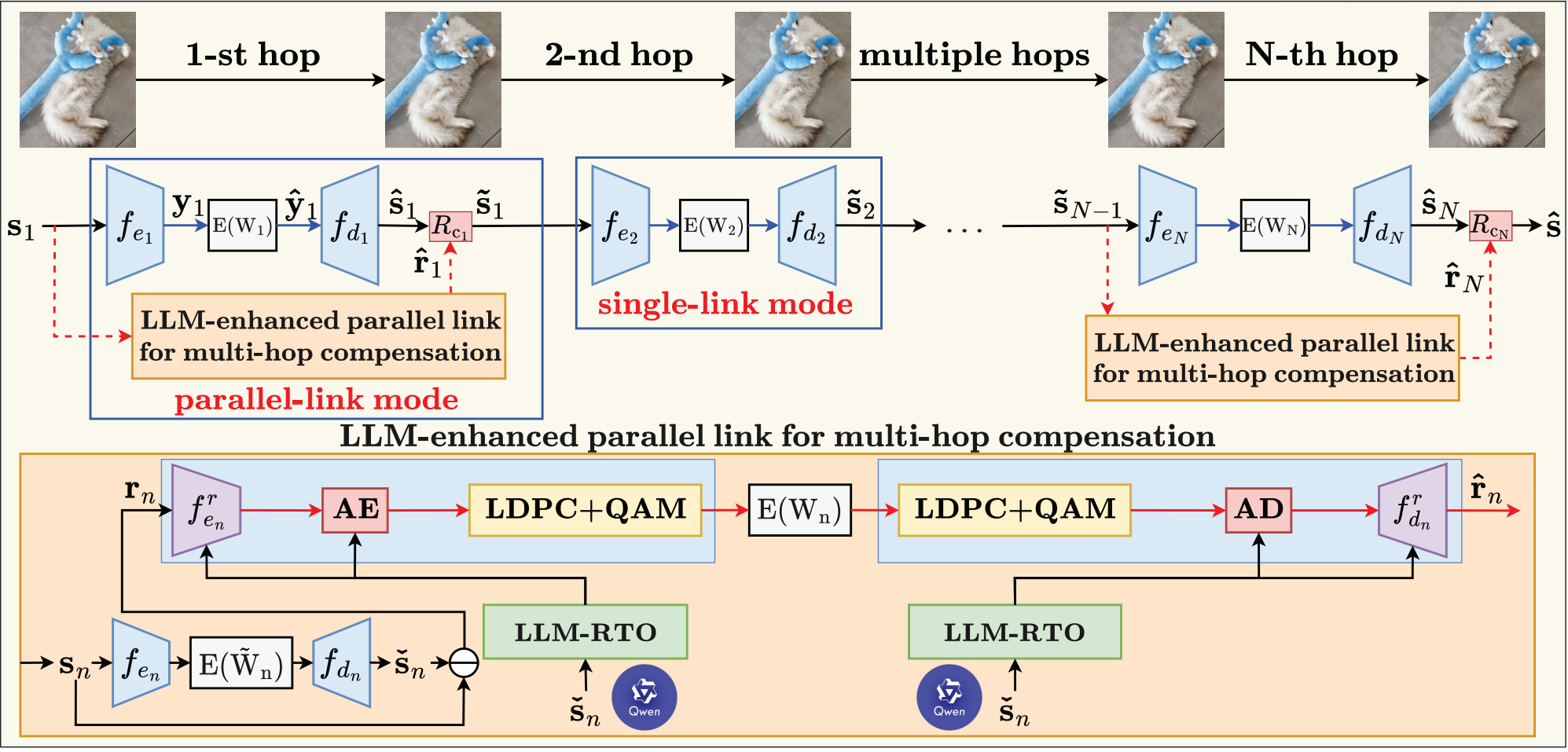}
	\caption{The proposed LLM-MHPSC framework based on a dual-link architecture, where LLM-enhanced residual links compensate for multi-hop error propagation over semantic links.}
	\label{fig_1}
\end{figure*}

\begin{enumerate}
	\item{}
	To address distortion accumulation in multi-hop wireless image transmission, we propose LLM-MHPSC, a novel dual-link framework that introduces an extra residual compensation link at each hop. By explicitly injecting compressed residual information to counteract error propagation, the proposed design effectively mitigates multi-hop performance degradation without altering the underlying single-hop semantic codec.
	\item{}
	To improve the efficiency of the residual compensation link, we develop a two-stage coarse-to-fine residual compression scheme. More importantly, we introduce an LLM-enhanced residual transmission optimizer (LLM-RTO), where a pretrained and partially fine-tuned LLM jointly processes reference image embeddings, residual embeddings, and CSI and hop-related side information embeddings. Through heterogeneous token interaction and global contextual modeling, LLM-RTO provides accurate residual distribution estimation for AAC and enables CSI and hop-aware residual rate adjustment, significantly enhancing compression efficiency with marginal extra bandwidth overhead.
	\item{}
	Since activating the residual compensation link at every hop introduces additional computational overhead, it is necessary to selectively compensate only the hops that are more critical to distortion mitigation. To this end, we propose a principled hop-aware compensation strategy with dynamically learned weighting coefficients. A lightweight gate network generates adaptive coefficients based on causal side information. This design enables data-driven hop-aware compensation and achieves an adaptive trade-off between reconstruction quality and latency.
	\item{}
	To validate the effectiveness of LLM-MHPSC, we conduct extensive experiments against state-of-the-art DL-based semantic communication schemes \cite{mhpsc, ADJSCC, mhdeepsc, WITT} as well as traditional SSCC methods. Results demonstrate that LLM-MHPSC consistently achieves superior reconstruction quality and robustness in multi-hop scenarios, with only a modest increase in bandwidth and computational overhead.
\end{enumerate}

Building upon the conference version \cite{mhpsc}, which established the MHPSC framework with coarse-to-fine residual compression, we make two key advancements. First, we develop a fine-tuned LLM-RTO to replace the original CNN-based residual estimator, enabling more accurate residual distribution modeling for AAC. Moreover, LLM-RTO incorporates side information, including CSI and hop index, to support joint residual rate adjustment and further reduce bandwidth overhead. Second, instead of activating the compensation link at all hops \cite{mhpsc}, we propose an adaptive hop selection strategy for on-demand residual compensation. This design improves the trade-off between reconstruction quality and computational cost, while enabling plug-and-play integration with existing single-link semantic communication frameworks.

Notations: $\mathbb{R}$ and $\mathbb{C}$ are the real and complex number sets. $\mathcal{CN}\left (\mu, \sigma^2 \right)$ denotes a complex Gaussian distribution with mean $\mu$ and variance $\sigma^2$. $\mathrm{diag}(\cdot)$ refers to the diagonalization operation between a vector and its corresponding diagonal matrix. $|\cdot|$ refers to computing the modulus of a complex number. $(\cdot)^*$ denotes complex conjugation. $\mathbf{I}$ denotes the unit matrix. The operator $\left(\cdot\right)^{T}$ denotes the matrix transpose.

\section{Overview of Proposed Framework}
In this section, we describe the multi-hop wireless image transmission framework with residual compensation link.

\subsection{Transmitter of LLM-MHPSC}

We consider a multi-hop wireless image transmission scenario, where images are forwarded across multiple wireless nodes. As shown in Fig. \ref{fig_1}, LLM-MHPSC adopts a dual-link architecture, consisting of a common semantic link and an LLM-enhanced residual compensation link. For the semantic link, marked by blue lines, the encoder $f_{e_n}(\cdot): \mathbb{R}^{H\times W\times 3}\rightarrow \mathbb{R}^{L}$ maps the input image $\mathbf{s}_n$ at the $n$-th hop to a codeword $\mathbf{y}_n \in \mathbb{R}^{L}$ for wireless transmission.

In parallel, the residual compensation link, marked by dashed red lines, mitigates accumulated distortion. Specifically, an emulated semantic transmission process with channel $\tilde{W}_n(\cdot)$ first produces a reference reconstruction $\check{\mathbf{s}}_n$, and the residual is computed as $\mathbf{r}_n=\mathbf{s}_n-\check{\mathbf{s}}_n$. The residual encoder $f_{e_n}^r(\cdot): \mathbb{R}^{H\times W\times 3}\rightarrow \mathbb{R}^{L^c\times L^c\times 3}$ compresses $\mathbf{r}_n$ into $\tilde{\mathbf{r}}_n$. Then, AAC is applied with residual distribution and rate allocation provided by LLM-RTO. The encoded residual bitstream is further processed by channel coding and modulation (LDPC+QAM) and mapped to complex symbols $\mathbf{r}_n^p \in \mathbb{C}^{L^r}$, where AE and AD denote the arithmetic encoder-decoder pair.

Notably, as illustrated by the two modes in Fig.~\ref{fig_1}, the compensation link is activated on demand rather than at every hop. The hop-aware selection strategy is detailed in Sec. III-D.

\subsection{Wireless Transmission}
For the semantic link, the codewords $\mathbf{y}_n\in\mathbb{C}^{L}$ is transmitted through a Rayleigh fading channel and then processed by minimum mean square error (MMSE) equalization as
\begin{align}
	\hat{\mathbf{y}}_n
	=
	\mathbf{G}_n\mathbf{y}_n
	+
	\mathbf{B}_n\mathbf{n}_n,
\end{align}
\begin{align}
	\mathbf{G}_n
	=
	\mathrm{diag}
	\left(
	\frac{|\mathbf{h}_n|^2}
	{|\mathbf{h}_n|^2+\sigma_n^2}
	\right),
	\mathbf{B}_n
	=
	\mathrm{diag}
	\left(
	\frac{\mathbf{h}_n^{*}}
	{|\mathbf{h}_n|^2+\sigma_n^2}
	\right),
\end{align}
where $\mathbf{h}_n=[h_{n,1},\ldots,h_{n,L}]^T\in\mathbb{C}^{L}$ denotes the Rayleigh fading vector with each element following $\mathcal{CN}(0,1)$, $\mathbf{n}_n\sim\mathcal{CN}(\mathbf{0},\sigma_n^2\mathbf{I})$ is the complex Gaussian noise vector.

Similarly, for the residual compensation link, the received residual symbols $\hat{\mathbf{r}}_n^p\in\mathbb{C}^{L^r}$ can be given as
\begin{align}
	\hat{\mathbf{r}}_n^p
	=
	\mathbf{G}_n^r\mathbf{r}_n^p
	+
	\mathbf{B}_n^r\mathbf{n}_n^r,
\end{align}
\begin{align}
	\mathbf{G}_n^r
	=
	\mathrm{diag}
	\left(
	\frac{|\mathbf{h}_n^r|^2}
	{|\mathbf{h}_n^r|^2+(\sigma_n^r)^2}
	\right),
	\mathbf{B}_n^r
	=
	\mathrm{diag}
	\left(
	\frac{(\mathbf{h}_n^r)^{*}}
	{|\mathbf{h}_n^r|^2+(\sigma_n^r)^2}
	\right),
\end{align}
where $\mathbf{h}_n^r\in\mathbb{C}^{L^r}$ and $\mathbf{n}_n^r$ are similar to definitions in Eq. (2).

\subsection{Receiver of LLM-MHPSC}

At the receiver, the semantic and residual links are jointly utilized for image reconstruction. For the semantic link, after complex-to-real conversion, the received codewords $\hat{\mathbf{y}}_n$ are decoded by $f_{d_n}(\cdot): \mathbb{R}^{L}\rightarrow \mathbb{R}^{H\times W\times 3}$ to obtain the reconstructed image $\hat{\mathbf{s}}_n$. 

Meanwhile, for the residual link, the received symbols $\hat{\mathbf{r}}_n^p$ are processed through QAM demodulation and SSCC decoding, followed by the residual decompressor $f_{d_n}^r(\cdot): \mathbb{R}^{L^c\times L^c\times3}\rightarrow \mathbb{R}^{H\times W\times 3}$ to recover the residual $\hat{\mathbf{r}}_n$. The final reconstructed image $\tilde{\mathbf{s}}_n$ is then obtained as
\begin{align}
	\tilde{\mathbf{s}}_n=(1-T_n)\hat{\mathbf{s}}_n+T_nR_{c_n}(\hat{\mathbf{s}}_n,\hat{\mathbf{r}}_n), T_n \in \left\{0,1\right\},
\end{align}
where $T_n=1$ activates the residual compensation link, while $T_n=0$ indicates semantic link-only transmission. $R_{c_n}(\cdot,\cdot)$ fuses the image and residuals for the refined $\tilde{\mathbf{s}}_n$.

For multi-hop transmission, the process starts from the source image $\mathbf{s}_1 \in \mathbb{R}^{H\times W\times 3}$. At the $n$-th hop, the input image is given by the reconstructed output from the previous hop, i.e., $\mathbf{s}_n = \tilde{\mathbf{s}}_{n-1}$. After $N$ successive hops, the final reconstructed image is obtained as $\hat{\mathbf{s}} = \tilde{\mathbf{s}}_N$ at the destination node. All hops share identical network structures and parameters as in \cite{mhdeepsc}. By integrating the residual compensation link in parallel, the proposed framework progressively corrects distortion along the transmission chain, effectively alleviating multi-hop error accumulation.

\section{Details of LLM-MHPSC}

\begin{figure*}[htbp]
	\centering
	\includegraphics[width=6.2in]{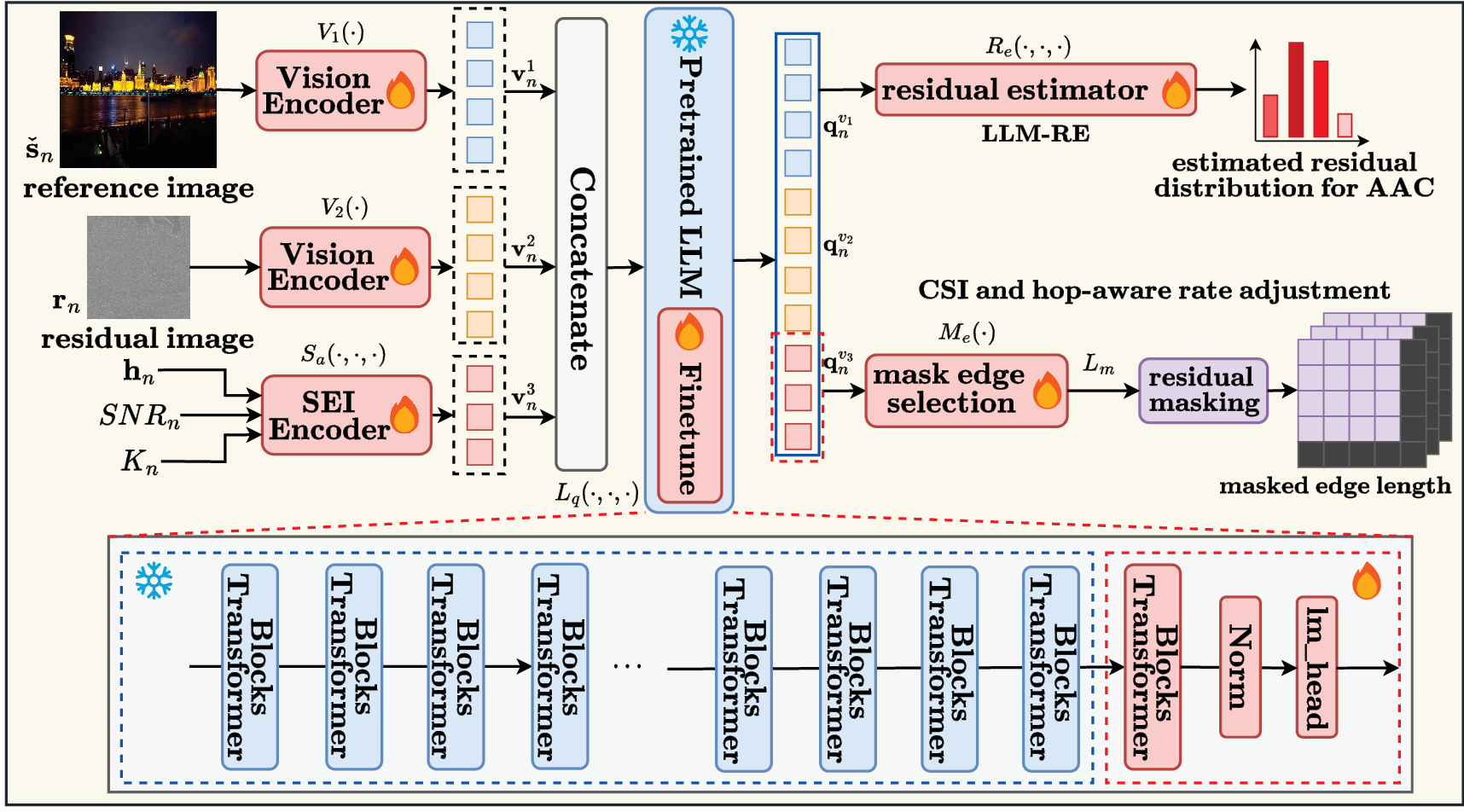}
	\caption{Illustration of the LLM-enhanced residual link transmission optimizer (LLM-RTO), which integrates LLM-based residual estimation and CSI and hop-aware rate adjustment.}
	\label{fig_2}
\end{figure*}

\subsection{Feature Extraction of LLM}
As illustrated in Fig.~\ref{fig_2}, LLM-RTO relies on heterogeneous information from both the visual and transmission domains. It should be emphasized that the proposed method is promoted by LLM not because the LLM generates textual outputs, but because the pretrained LLM is used as a contextual optimizer for multi-source token embeddings. Specifically, reference-image embeddings, residual-image embeddings, and supplemental enhancement information (SEI) embeddings are projected into the LLM-compatible space and jointly processed by the partially fine-tuned LLM. Through self-attention, the LLM captures the dependencies among image semantics, residual structures, channel states, and hop-dependent transmission conditions, which are essential for residual distribution estimation and residual-link rate control.

Specifically, the reference image $\check{\mathbf{s}}_n$ and residual image $\mathbf{r}_n$ are mapped into visual embeddings as
\begin{align}
	\mathbf{v}_n^{1} = V_1(\check{\mathbf{s}}_n),
\end{align}
\vspace{-20pt}
\begin{align}
	\mathbf{v}_n^{2} = V_2(\mathbf{r}_n),
\end{align}
where $V_1(\cdot): \mathbb{R}^{H\times W\times 3}\rightarrow \mathbb{R}^{L_{v_1}\times D}$ and $V_2(\cdot): \mathbb{R}^{H\times W\times 3}\rightarrow \mathbb{R}^{L_{v_2}\times D}$ denote the vision encoders. They transform image-domain inputs into token embeddings $\mathbf{v}_n^{1}$ and $\mathbf{v}_n^{2}$ aligned with the LLM embedding space. Here, $L_{v_1}$ and $L_{v_2}$ denote the corresponding token lengths, and $D$ is the hidden dimension of the LLM.

In addition to visual inputs, SEI is introduced to provide transmission-aware context for multi-hop residual optimization. Specifically, the channel fading coefficient $\mathbf{h}_n$, signal-to-noise ratio $\mathrm{SNR}_n$, and hop-related indicator $K_n$ are encoded by a dedicated SEI encoder as
\begin{align}
	\mathbf{v}_n^{3} = S_a(\mathbf{h}_n, \mathrm{SNR}_n, K_n),
\end{align}
where $S_a(\cdot,\cdot,\cdot)$ maps channel and hop-related descriptors into $\mathbf{v}_n^{3} \in \mathbb{R}^{L_{v_3}\times D}$. These SEI embeddings are essential because residual statistics in multi-hop transmission are affected not only by image content, but also by channel variation, hop dependency, and accumulated distortion.

The embeddings $\mathbf{v}_n^{1}$, $\mathbf{v}_n^{2}$, and $\mathbf{v}_n^{3}$ are then concatenated and jointly processed by a partially fine-tuned LLM:
\begin{align}
	\mathbf{q}^{v_1}_n,\mathbf{q}^{v_2}_n,\mathbf{q}^{v_3}_n
	=
	L_q(\mathbf{v}_n^{1}, \mathbf{v}_n^{2}, \mathbf{v}_n^{3}),
\end{align}
where $L_q(\cdot,\cdot,\cdot)$ denotes the pretrained LLM. The output features $\mathbf{q}^{v_1}_n\in \mathbb{R}^{L_{v_1}\times D}$, $\mathbf{q}^{v_2}_n\in \mathbb{R}^{L_{v_2}\times D}$, and $\mathbf{q}^{v_3}_n\in \mathbb{R}^{L_{v_3}\times D}$ are context-enhanced representations generated through joint visual-SEI modeling.

The extracted features play complementary roles in the following modules. The visual features $\mathbf{q}^{v_1}_n$ and $\mathbf{q}^{v_2}_n$ provide semantic structure and fine-grained residual cues for residual distribution estimation in AAC. Meanwhile, the SEI-related feature $\mathbf{q}^{v_3}_n$ captures channel-state and hop-aware transmission characteristics for inter-hop residual rate adjustment. Therefore, the benefit of LLM-RTO comes from the synergy between heterogeneous feature fusion and contextual dependency modeling: visual and SEI embeddings provide the necessary multi-source information, while the LLM further exploits their correlations for accurate residual compression and efficient multi-hop transmission control.

\subsection{LLM for Residual Estimation}

As illustrated in Fig. \ref{fig_2}, the LLM-enhanced residual estimator (LLM-RE) is designed to predict the probability distribution of compressed residuals for adaptive arithmetic coding (AAC). Since AAC relies on an accurate source distribution to approach efficient lossless compression, precise residual distribution estimation is essential for reducing the overhead of the compensation link. However, in multi-hop transmission, residual statistics are not only determined by local image content, but also affected by accumulated distortion, channel variation, and hop-dependent transmission conditions. This motivates the use of LLM-extracted features \cite{deletang, llmrc, next-pixel, nmi} to provide a richer contextual representation for residual modeling.

Based on the LLM feature extraction, the reference image, residual image, and SEI are first transformed into LLM-compatible embeddings and jointly processed by the partially fine-tuned LLM. The resulting features $\mathbf{q}^{v_1}_n$, $\mathbf{q}^{v_2}_n$, and $\mathbf{q}^{v_3}_n$ contain complementary information for residual estimation. By jointly exploiting these features, LLM-RE can estimate residual distributions in a semantic-aware and transmission-aware manner.

Given the LLM output features, the residual estimator $R_e(\cdot,\cdot,\cdot)$ predicts the probability mass function of $\tilde{\mathbf{r}}_n$ as
\begin{align}
	p(\tilde{\mathbf{r}}_n|\mathbf{q}^{v_1}_n,\mathbf{q}^{v_2}_n,\mathbf{q}^{v_3}_n)
	=
	R_e(\mathbf{q}^{v_1}_n,\mathbf{q}^{v_2}_n,\mathbf{q}^{v_3}_n).
\end{align}

Following \cite{rc}, a discrete mixture of logistic distributions is adopted to model $\tilde{\mathbf{r}}_n$. Let $c=1,2,3$ denote the RGB channels, and let $(u,v)$ denote the spatial location. The likelihood of $\tilde{\mathbf{r}}_n$ can be factorized over spatial positions as
\begin{align}
	p(\tilde{\mathbf{r}}_n|\mathbf{q}^{v_1}_n,\mathbf{q}^{v_2}_n,\mathbf{q}^{v_3}_n)
	=
	\prod_{u,v}
	p(\tilde{\mathbf{r}}_n^{1uv},\tilde{\mathbf{r}}_n^{2uv},\tilde{\mathbf{r}}_n^{3uv}
	|\mathbf{q}^{v_1}_n,\mathbf{q}^{v_2}_n,\mathbf{q}^{v_3}_n).
\end{align}

To capture the correlation among RGB channels with low complexity, we employ a weak autoregressive formulation over channels as
\begin{equation}
	\begin{aligned}
		& p(\tilde{\mathbf{r}}_n^{1uv},\tilde{\mathbf{r}}_n^{2uv},\tilde{\mathbf{r}}_n^{3uv}
		|\mathbf{q}^{v_1}_n,\mathbf{q}^{v_2}_n,\mathbf{q}^{v_3}_n)  \\
		& =
		p_m(\tilde{\mathbf{r}}_n^{1uv}|\mathbf{q}^{v_1}_n,\mathbf{q}^{v_2}_n,\mathbf{q}^{v_3}_n)
		\cdot
		p_m(\tilde{\mathbf{r}}_n^{2uv}|\mathbf{q}^{v_1}_n,\mathbf{q}^{v_2}_n,\mathbf{q}^{v_3}_n,\tilde{\mathbf{r}}_n^{1uv}) \\
		& \quad \cdot
		p_m(\tilde{\mathbf{r}}_n^{3uv}|\mathbf{q}^{v_1}_n,\mathbf{q}^{v_2}_n,\mathbf{q}^{v_3}_n,
		\tilde{\mathbf{r}}_n^{1uv},\tilde{\mathbf{r}}_n^{2uv}).
	\end{aligned}
\end{equation}

The mixture distribution $p_m$ is represented by a mixture of $K$ logistic distributions. The residual estimator outputs the mixture weights $\pi_n^{k,cuv}$, means $\mu_n^{k,cuv}$, variances $\sigma_n^{k,cuv}$, and channel-wise mixture coefficients $\lambda_n^{k,cuv}$. The autoregressive dependency is introduced by adjusting the means according to previously decoded channels:
\begin{align}
	\tilde{\mu}_n^{k,cuv}
	=
	\begin{cases}
		\mu_n^{k,1uv}, & c=1,\\
		\mu_n^{k,2uv}+\lambda_n^{k,1uv}\tilde{\mathbf{r}}_n^{1uv}, & c=2,\\
		\mu_n^{k,3uv}+\lambda_n^{k,2uv}\tilde{\mathbf{r}}_n^{1uv}
		+\lambda_n^{k,3uv}\tilde{\mathbf{r}}_n^{2uv}, & c=3.
	\end{cases}
\end{align}

Thus, the mixture probability is formulated as
\begin{equation}
	\begin{aligned}
		& p_m\!\left(
		\tilde{\mathbf{r}}_n^{cuv}
		\mid
		\mathbf{q}^{v_1}_n,\mathbf{q}^{v_2}_n,\mathbf{q}^{v_3}_n,
		\tilde{\mathbf{r}}_n^{\mathrm{prev}}
		\right)  \\
		&\quad =
		\sum_{k=1}^{K}
		\pi_n^{k,cuv}
		p_L\!\left(
		\tilde{\mathbf{r}}_n^{cuv}
		\mid
		\tilde{\mu}_n^{k,cuv},
		\sigma_n^{k,cuv}
		\right),
	\end{aligned}
\end{equation}
where $\tilde{\mathbf{r}}_n^{\mathrm{prev}}$ denotes the channels before channel $c$.

The logistic distribution is given by
\begin{align}
	p_L(\tilde{\mathbf{r}}_n|\tilde{\mu}_n^{k,cuv},\sigma_n^{k,cuv})
	=
	\frac{
		e^{-(\tilde{\mathbf{r}}_n-\tilde{\mu}_n^{k,cuv})/\sigma_n^{k,cuv}}
	}{
		\sigma_n^{k,cuv}
		\left(
		1+
		e^{-(\tilde{\mathbf{r}}_n-\tilde{\mu}_n^{k,cuv})/\sigma_n^{k,cuv}}
		\right)^2
	}.
\end{align}

Since the compressed residual is discrete, the probability mass $p_L$ is obtained by evaluating the cumulative distribution function (CDF) as
\begin{align}
	p_L(\tilde{\mathbf{r}}_n)
	=
	\mathrm{CDF}\left(\tilde{\mathbf{r}}_n+\frac{1}{2}\right)
	-
	\mathrm{CDF}\left(\tilde{\mathbf{r}}_n-\frac{1}{2}\right).
\end{align}

\begin{figure}[htbp]
	\centering
	\includegraphics[width=3.0in]{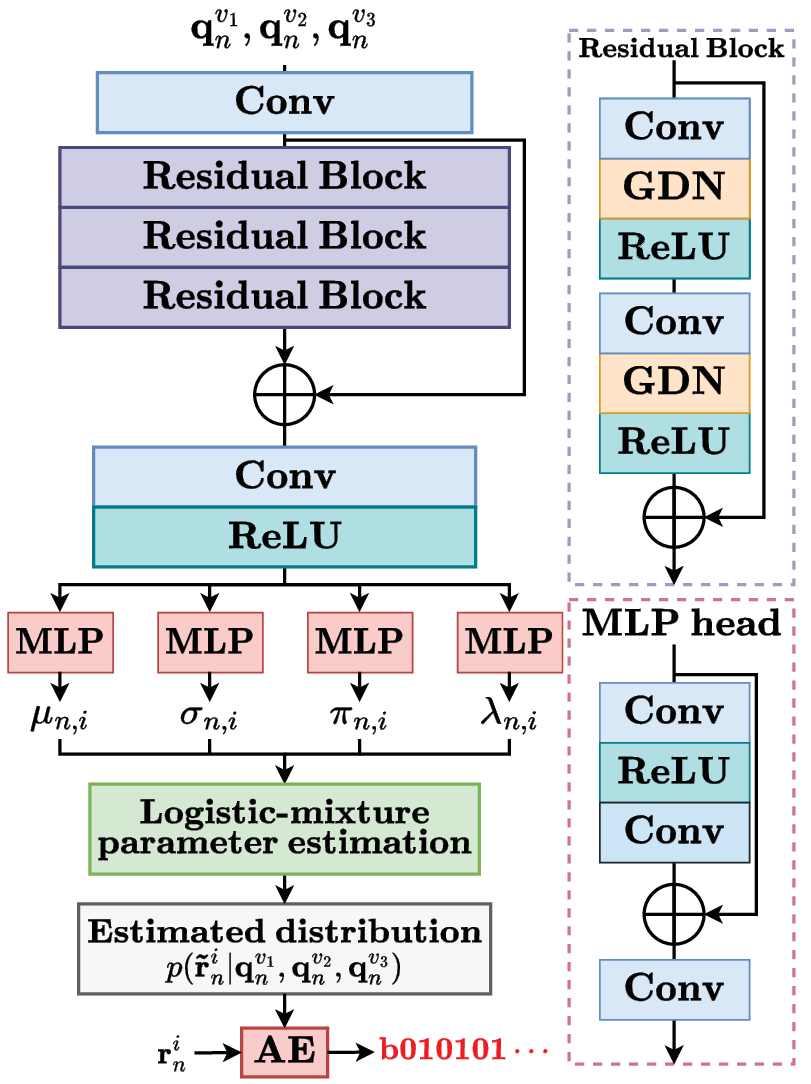}
	\caption{The structure of LLM-RE.}
	\label{fig_3}
\end{figure}

As shown in Fig. \ref{fig_3}, the residual estimator is implemented by stacked residual blocks followed by multi-layer perceptron (MLP) heads. These heads estimate the parameters $\mu_{n,i}$, $\sigma_{n,i}$, $\pi_{n,i}$, and $\lambda_{n,i}$ required by the learned distributions. Compared with a CNN-based residual estimator \cite{mhpsc}, the LLM-enhanced estimator has two advantages. First, CNNs mainly capture local spatial dependencies, while the LLM can model global correlations among visual and transmission-related tokens. Second, the residual distribution in multi-hop transmission is not only determined by image content alone, but also depends on channel states and hop-dependent accumulated distortion. By incorporating SEI embeddings into the LLM token sequence, the proposed estimator becomes both semantic and transmission-aware, leading to more compression-efficient probability modeling for AAC.

\subsection{LLM for CSI and Hop-aware Rate Adjustment}

The LLM-enhanced residual estimator has incorporated visual features and SEI features for hop-aware residual distribution estimation. However, accurate distribution estimation alone does not explicitly regulate how much residual information should be transmitted at each hop. In multi-hop transmission with a fixed average CBR budget, different hops may require different residual rates. For example, hops with severe channel fading or larger accumulated distortion should preserve more residual information, while hops with favorable channel conditions can tolerate stronger compression. Therefore, beyond residual distribution estimation, it is necessary to further perform inter-hop residual-link rate adjustment.

To this end, LLM-RTO further exploits the SEI-related LLM output feature $\mathbf{q}^{v_3}_n$ to control the residual transmission rate. As shown in Fig.~\ref{fig_2}, $\mathbf{q}^{v_3}_n$ encodes CSI and hop-related information, which provides a compact representation for determining the compensation intensity of the $n$-th hop. Based on this feature, the mask edge selector $M_e(\cdot)$ predicts the appropriate masking edge length as
\begin{align}
	L^m = \arg\max\left(M_e(\mathbf{q}^{v_3}_n)\right).
\end{align}

The corresponding masking ratio is computed by
\begin{align}
	r_m = 1-\left(1-\frac{L^m}{L^c}\right)^2,
\end{align}
where $L^c$ is the spatial size of the compressed residual representation. A larger $L^m$ indicates that more boundary regions of the compressed residual are masked, resulting in a lower residual transmission rate.

The masked residual representation is then obtained as
\begin{align}
	\mathbf{r}'_n = M_a(\tilde{\mathbf{r}}_n,L^m),
\end{align}
where $M_a(\cdot,\cdot)$ denotes the residual masking operation. Through this mechanism, LLM-RTO adaptively adjusts the residual link rate among different hops. Critical hops can retain more residual information for distortion mitigation, while less critical hops transmit fewer residual symbols to reduce bandwidth overhead. This enables CSI and hop-aware residual rate allocation under the overall CBR constraint.

Since the pretrained LLM is not originally optimized for residual compression and inter-hop rate control, direct deployment may lead to task mismatch. Therefore, we adopt a partial finetuning (FT) strategy to adapt LLM-RTO while controlling training cost \cite{lora,qlora}. As shown in Fig.~\ref{fig_2}, most Transformer blocks are frozen, and only the last Transformer block, the normalization layer, and the language modeling head (\texttt{lm\_head}) are finetuned. Meanwhile, the vision encoders, SEI encoder, mask edge selector, and residual estimator are jointly optimized with the finetuned LLM components. This preserves the contextual modeling ability of the pretrained LLM while enabling task-specific adaptation for hop-aware residual distribution estimation and inter-hop rate adjustment.

\subsection{Latency and Quality Trade-off Transmission Strategy among Multiple Hops}

Although the LLM-enhanced residual compensation link effectively mitigates distortion accumulation, activating it at every hop introduces additional computational overhead. Therefore, it is unnecessary and inefficient to compensate all hops uniformly. In practical multi-hop transmission, only hops with severe channel degradation, accumulated distortion risk, or insufficient compensation history are critical for compensation. Since LLM-MHPSC augments the semantic link with an optional residual compensation path, deactivating the residual link does not affect the basic semantic transmission. This enables flexible on-demand compensation and provides a practical trade-off between reconstruction quality and latency.

The activation of the residual compensation link is determined by three factors: the current channel condition, the accumulated channel degradation since the last compensated hop, and the remaining compensation budget. We use the channel capacity as a unified metric for link quality. For the $n$-th hop, the average channel capacity $C_n$ is given by
\begin{align}
	C_n
	=
	\frac{1}{L}
	\sum_{l=1}^{L}
	\log_2
	\left(
	1+
	|h_{n,l}|^2\mathrm{SNR}_n
	\right).
\end{align}

Based on the above factors, the compensation probability is formulated as
\begin{align}
	P_n=\min\left(
	\lambda_{1,n}\frac{C_{\mathrm{th}}}{C_n}
	+
	\lambda_{2,n}\sum_{k=n_c}^{n}\frac{C_{\mathrm{th}}}{C_k}
	+
	\lambda_{3,n}\frac{N_c-N_t}{N_c},
	1
	\right),
\end{align}
where $C_{\mathrm{th}}$ is the channel capacity threshold, $n_c$ denotes the latest compensated hop, $N_c$ is the compensated hop budget, and $N_t$ is the compensated hop count used so far. $\lambda_{1,n}$, $\lambda_{2,n}$, and $\lambda_{3,n}$ respectively measure the instantaneous channel degradation, the historical accumulated degradation, and the remaining compensation budget.

\begin{figure}[htbp]
	\centering
	\includegraphics[width=3.4in]{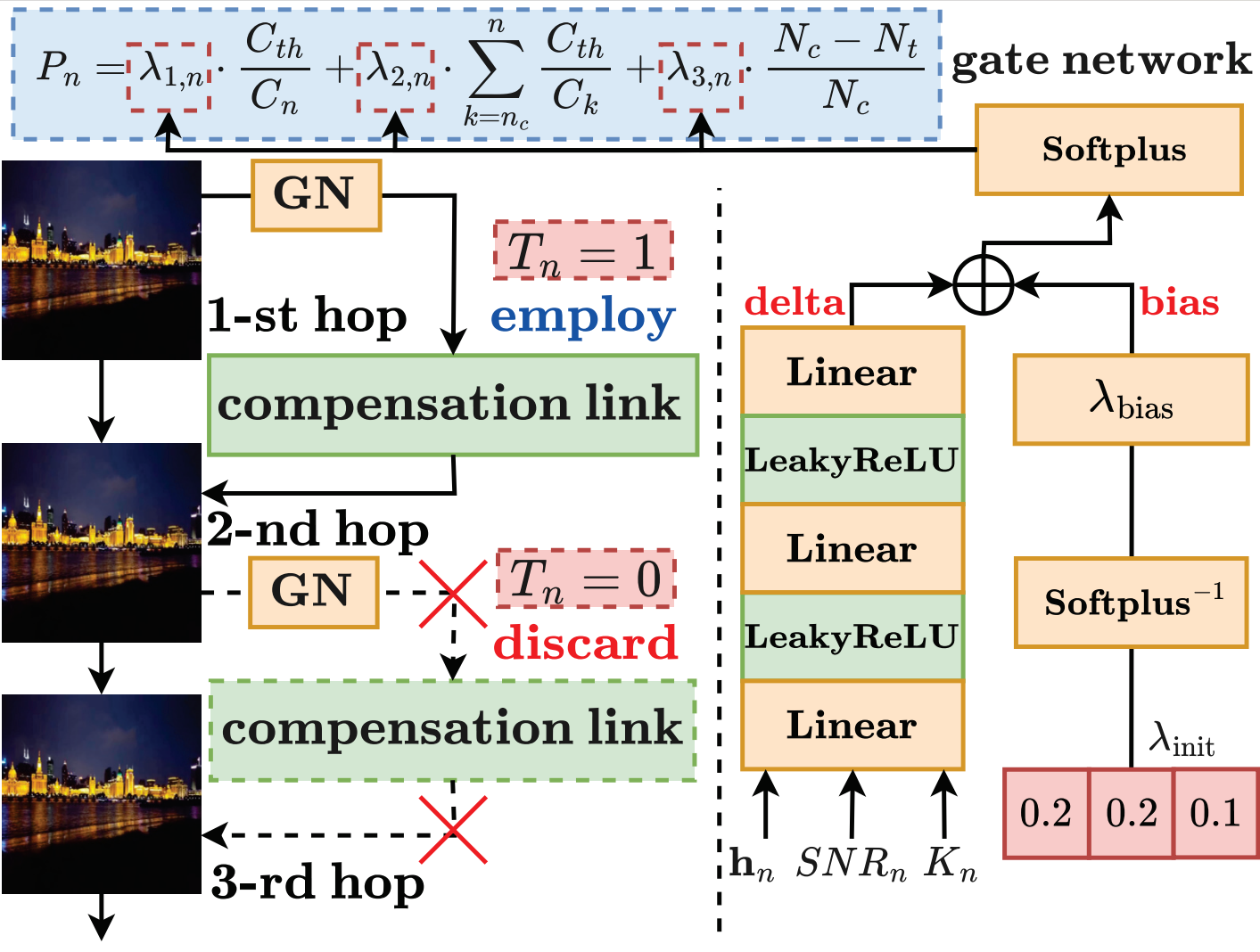}
	\caption{Illustration of the proposed adaptive compensation link selection strategy with a lightweight gate network.}
	\label{fig_4}
\end{figure}

As illustrated in Fig.~\ref{fig_4}, the binary compensation decision is obtained by comparing $P_n$ with a threshold $P'$:
\begin{align}
	T_n = 
	\begin{cases}
		1, & n=1,\\
		1, & P_n \ge P',\\
		0, & P_n < P',
	\end{cases}
\end{align}
where $T_n=1$ indicates that the residual compensation link is activated, while $T_n=0$ indicates that only the semantic link is used. The first hop is always compensated because its reconstruction quality affects all subsequent hops.

To avoid manually fixed weighting coefficients in Eq. (21), a lightweight gate network (GN), $\Lambda(\cdot,\cdot,\cdot)$, is introduced to generate adaptive coefficients for each hop. As shown in Fig.~\ref{fig_4}, the GN takes hop-dependent SEI as input, including the channel fading coefficient $\mathbf{h}_n$, $\mathrm{SNR}_n$, and hop-related indicator $K_n$. It first extracts a compact latent representation and then predicts a residual adjustment term $\Delta\boldsymbol{\lambda}_n$.  

Meanwhile, a learnable bias term $\boldsymbol{\lambda}_{\mathrm{bias}}$ is initialized from predefined coefficients $\boldsymbol{\lambda}_{\mathrm{init}}=(0.2,0.2,0.1)$ through the inverse Softplus transformation, $\mathrm{Softplus}^{-1}(\cdot)$. Then, $\Delta\boldsymbol{\lambda}_n$ is combined with learnable bias parameters and passed through a Softplus activation as
\begin{align}
	\boldsymbol{\lambda}_n
	=
	\mathrm{Softplus}
	\left(
	\boldsymbol{\lambda}_{\mathrm{bias}}
	+
	\Delta\boldsymbol{\lambda}_n
	\right),
\end{align}
where $\boldsymbol{\lambda}_n=[\lambda_{1,n},\lambda_{2,n},\lambda_{3,n}]$, $\boldsymbol{\lambda}_{\mathrm{bias}}=\mathrm{Softplus}^{-1}(\boldsymbol{\lambda}_{\mathrm{init}})$.

The proposed gate network enables the hop selection (HS) rule to adaptively balance different decision factors according to real-time transmission conditions. Compared with fixed-weight hop selection, it provides a data-driven compensation strategy. Moreover, GN only uses causal information available at the current and previous hops, without requiring future channel states, making it suitable for practical deployment.

Finally, we analyze the computation cost across multiple hops. Since the residual compensation link involves LLM-based processing, its computation cost is generally higher than that of the semantic link. The total computation time is formulated as
\begin{align}
	T_h=\sum_{n=1}^{N}
	\left[
	T_nT^c_n+(1-T_n)T^s_n
	\right],
\end{align}
where $T^c_n$ and $T^s_n$ denote the runtime of the compensation link and the semantic link at the $n$-th hop, respectively. By selectively activating the residual link, the proposed strategy reduces unnecessary LLM computation while preserving the quality gain of residual compensation.

\section{Deployment Detail}
In this section, we present the network structure and training strategy of the LLM-MHPSC framework.

\subsection{Network Structure}
\begin{figure}[htbp]
	\centering
	\includegraphics[width=3.2in]{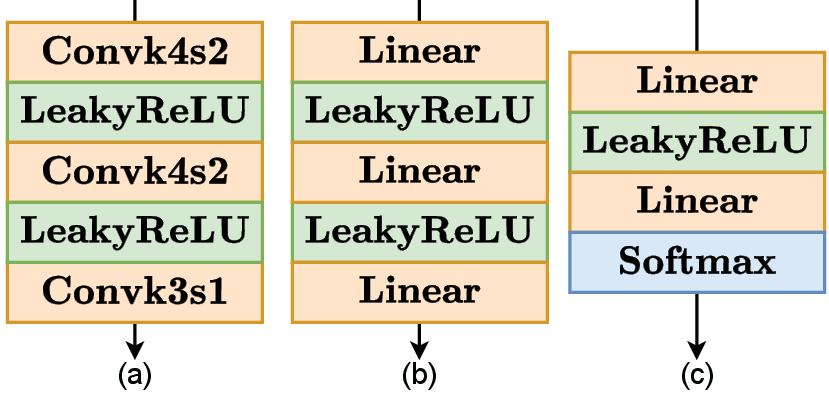}
	\caption{(a) Vision encoder (b) SEI encoder (c) Mask edge selector}
	\label{fig_5}
\end{figure}

Following Fig. \ref{fig_1}, the main semantic backbone is implemented based on the architecture in \cite{WITT}. The residual compressor is the same as the residual encoder and decoder in \cite{WVSC}. The vision encoders $V_1(\cdot)$ and $V_2(\cdot)$ are CNN-based structures while SEI encoder $S_a(\cdot,\cdot,\cdot)$ and mask edge selection module $M_e(\cdot)$ are deep neural network (DNN)-based structures, as shown in Fig. \ref{fig_5}. For the pretrained LLM $L_q(\cdot,\cdot,\cdot)$, Qwen3-1.7B \cite{qwen3} is adopted as a lightweight LLM for residual link transmission optimization. To further accelerate LLM deployment speed, we deploy LLM-MHPSC with BFloat16.

\begin{algorithm}[htbp]
	\caption{Finetuning for proposed LLM-RTO}\label{alg:alg1}
	\begin{algorithmic}
		\STATE 
		$\textbf{Input:}$ Raw image $\mathbf{s}_n$,  randomly sampled $\mathbf{h}_n,SNR_n,K_n$
		
		$\textbf{Output:}$ Well-trained LLM-RTO containing $V_1(\cdot)$, $V_2(\cdot)$, $S_a(\cdot,\cdot,\cdot)$, $M_e(\cdot)$, $L_q(\cdot,\cdot,\cdot)$, $R_e(\cdot,\cdot,\cdot)$
		
		\STATE1. Freeze the weights of $f_{e_n}(\cdot)$, $f_{d_n}(\cdot)$, $f_{e_n}^r(\cdot)$, $f_{d_n}^r(\cdot)$
		\STATE2. Extract the visual and SEI embeddings
		\STATE \hspace{0.25cm} $\mathbf{v}^1_n=V_1(\mathbf{\check{s}}_n), \mathbf{v}^2_n=V_2(\mathbf{r}_n), \mathbf{v}^3_n=S_a(\mathbf{h}_n,SNR_n,K_n)$
		\STATE3. Generate LLM output features
		\STATE \hspace{0.25cm} $\mathbf{q}^{v_1}_n,\mathbf{q}^{v_2}_n,\mathbf{q}^{v_3}_n=L_q(\mathbf{v}^{1}_n,\mathbf{v}^{2}_n,\mathbf{v}^{3}_n)$
		\STATE4. Estimate the residual distribution
		\STATE \hspace{0.25cm} $p(\mathbf{\tilde{r}}_n|\mathbf{q}^{v_1}_n,\mathbf{q}^{v_2}_n,\mathbf{q}^{v_3}_n)=R_e(\mathbf{q}^{v_1}_n,\mathbf{q}^{v_2}_n,\mathbf{q}^{v_3}_n)$
		\STATE5. Select mask edge length $L^m$ with Softmax operation
		\STATE6. Perform CSI and hop-aware rate adjustment
		\STATE \hspace{0.25cm} $\mathbf{r}'_n=M_a(\mathbf{\tilde{r}}_n, L^m)$
		\STATE7. Compute the loss $L_\mathrm{RTO}$ by Eq. (26)
		\STATE8. Update the finetuned network weights of $V_1(\cdot)$, $V_2(\cdot)$, $S_a(\cdot,\cdot,\cdot)$, $M_e(\cdot)$, $L_q(\cdot,\cdot,\cdot)$, $R_e(\cdot,\cdot,\cdot)$
		
		\STATE return the well-trained LLM-RTO		
	\end{algorithmic}
	\label{alg1}
\end{algorithm}

\begin{algorithm}[htbp]
	\caption{Training algorithm for the adaptive $\lambda$ selector}\label{alg:alg2}
	\begin{algorithmic}
		\STATE 
		$\textbf{Input:}$ Raw image $\mathbf{s}_n$,  randomly sampled $\mathbf{h}_n,SNR_n,K_n$
		
		$\textbf{Output:}$ Well-trained $\Lambda(\cdot,\cdot,\cdot)$
		
		\STATE1. Fix the weights of $f_{e_n}(\cdot)$, $f_{d_n}(\cdot)$, $f_{e_n}^r(\cdot)$, $f_{d_n}^r(\cdot)$
		\STATE2. Initialize the multi-hop parameters: $K = 0$
		\STATE3. For each training hop $n = 1,...,N$ do
		\STATE4. \hspace{0.25cm} $\lambda_1,\lambda_2,\lambda_3=\Lambda(\mathbf{h}_n,SNR_n,K_n)$
		\STATE5. \hspace{0.25cm} Compute the compensation probability $P_n$ by Eq. (21)
		\STATE6. \hspace{0.25cm} Decide the compensation choice $T_n$ by Eq. (22)
		\STATE7. \hspace{0.25cm} if $T_n=1$:
		
		\hspace{1.25cm} Employ residual link for hop compensation
		\STATE8. \hspace{0.25cm} else if $T_n=0$:
		
		\STATE \hspace{1.25cm} Utilize only semantic link for transmission
		\STATE9. Compute the loss $L_\lambda$ by Eq. (27)
		\STATE10. Update the finetuned network weights of $\Lambda(\cdot,\cdot,\cdot)$
		
		\STATE return the well-trained adaptive $\lambda$ selector		
	\end{algorithmic}
	\label{alg2}
\end{algorithm}

\begin{algorithm}[htbp]
	\caption{Training algorithm for proposed LLM-MHPSC}\label{alg:alg3}
	\begin{algorithmic}
		\STATE 
		$\textbf{Input:}$ Raw image $\mathbf{s}_n$, Number of hops N, randomly sampled mask edge length $L^m$	
		
		$\textbf{Output:}$ Well-trained $f_{e_n}(\cdot)$, $f_{d_n}(\cdot)$, $f_{e_n}^r(\cdot)$, $f_{d_n}^r(\cdot)$, LLM-RTO
		
		\STATE \textbf{Stage1:}
		\STATE1. For each training hop $n = 1,...,N$ do
		\STATE2. \hspace{0.5cm} $\mathbf{s}_n \xrightarrow{f_{e_n}(\cdot)} \mathbf{y}_n \xrightarrow{E\left(W_n(\cdot)\right)} \mathbf{\hat{y}}_n \xrightarrow{f_{d_n}(\cdot)} \mathbf{\hat{s}}_n$
		\STATE3. \hspace{0.5cm} Compute the loss $L_1$ by Eq. (25)
		\STATE4. \hspace{0.5cm} Update the network weights of $f_{e_n}(\cdot)$, $f_{d_n}(\cdot)$
		
		\STATE \textbf{Stage2:}
		\STATE5. For each training hop $n = 1,...,N$ do
		\STATE6. \hspace{0.5cm} Randomly sample $L^m$ for the current hop
		\STATE7. \hspace{0.5cm} $\mathbf{s}_n \xrightarrow{f_{e_n}(\cdot)} \mathbf{y}_n \xrightarrow{E\left(\tilde{W}_n(\cdot)\right)} \mathbf{\hat{y}}_n \xrightarrow{f_{d_n}(\cdot)} \mathbf{\check{s}}_n$
		\STATE8. \hspace{0.5cm} $\mathbf{r}_n=\mathbf{s}_n-\mathbf{\check{s}}_n$
		\STATE9. \hspace{0.5cm} $\mathbf{r}_n \xrightarrow{f_{e_n}^r(\cdot)} \mathbf{\tilde{r}}_n \xrightarrow{M_a(\cdot,L^m)} \mathbf{r}'_n\xrightarrow{f_{d_n}^r(\cdot)} \mathbf{\hat{r}}_n$
		\STATE10. \hspace{0.35cm} $\mathbf{{\tilde{s}}}_n=R_{c_n}(\mathbf{\hat{s}}_n,\mathbf{\hat{r}}_n)$
		\STATE11. Compute the loss $L_1$ by Eq. (25)
		\STATE12. Update the network weights of $f_{e_n}^r(\cdot)$, $f_{d_n}^r(\cdot)$
		
		\STATE \textbf{Stage3:}
		\STATE13. Finetune the network weights of LLM-RTO according to Alg. 1

		\STATE \textbf{Stage4:}
		\STATE14. Train the adaptive $\lambda$ selector and add hop selection strategy according to Alg. 2

		\STATE return the well-trained LLM-MHPSC		
	\end{algorithmic}
	\label{alg3}
\end{algorithm}

\subsection{Training Loss and Strategy}
The proposed LLM-MHPSC is trained in four stages by combining multi-hop recursive training with single-hop residual optimization. For the first stage, we train solely the network backbone $\mathbf{f}_{e_n}(\cdot)$ and $\mathbf{f}_{d_n}(\cdot)$ with multiple hops enhanced by the recursive training strategy in \cite{mhdeepsc} as
\begin{align}
	L_\mathrm{1}=\frac{1}{N\times I}\sum_{i=1}^{I}\sum_{n=1}^{N}(\gamma^{N-n}D\left(\mathbf{s}_n^i,\mathbf{\hat{s}}_n^i\right)),
\end{align}
where $\gamma$ refers to the scaling factor for adjusting the weight of each hop. $I$ is the total image number. $D(\cdot,\cdot)$ refers to the loss term which is set as mean square error (MSE) in default.

For the second stage, the parameters of the semantic transmission backbone are frozen, while the residual compressor $\mathbf{f}_{e_n}^r(\cdot)$ and decompressor $\mathbf{f}_{d_n}^r(\cdot)$ are introduced into the training pipeline. To accommodate the flexible rate adjustment enabled by the subsequently integrated LLM-RTO, the masking edge length $L^m$ is randomly sampled during this stage. The training objective remains the same as Eq. (25).

For the third stage, LLM-RTO-related modules, $V_1(\cdot)$, $V_2(\cdot)$, $S_a(\cdot,\cdot,\cdot)$, $M_e(\cdot)$, $L_q(\cdot,\cdot,\cdot)$, $R_e(\cdot,\cdot,\cdot)$ are finally added into the network training process to provide accurate residual distribution along with residual link rate adjustment to the AAC. We use Softmax to select $L_m$ during training while employing Eq. (17) for inference. To reduce training cost, we adopt a partial FT strategy, where only the last Transformer layer, the normalization layer, and the $\texttt{lm\_head}$ are updated, while the remaining LLM backbone is frozen. LLM is finetuned and trained along with other modules in LLM-RTO to provide residual link optimization. The LLM-RTO finetuning algorithm is summarized in Alg. 1. The training loss is given as
\begin{align}
	L_\mathrm{RTO}=-\frac{1}{I}\sum_{i=1}^{I}\log p(\mathbf{\tilde{r}}_n^i|\mathbf{\check{s}}_n^i,\mathbf{r}_n^i).
\end{align}

Finally, to accurately perform the multi-hop selection rather than an arbitrary manner, we train the adaptive $\lambda$ selector $\Lambda(\cdot,\cdot,\cdot)$ in Alg. 2. Only $\Lambda(\cdot,\cdot,\cdot)$ is activated for the whole network and the training loss is formulated as 
\begin{align}
	L_\lambda=\frac{1}{I}\sum_{i=1}^{I}D\left(\mathbf{s}^i,\mathbf{\hat{s}}^i\right)+\alpha\left(\frac{1}{N\times I}\sum_{i=1}^{I}\sum_{n=1}^{N} P_n-\frac{N_c}{N}\right),
\end{align}
where the first term measures the final reconstruction distortion, and the second term regulates the expected compensation ratio according to the predefined compensation budget $N_c/N$. The coefficient $\alpha$ balances reconstruction quality and latency-related compensation overhead.

Since LLM-RTO mainly aims to reduce the CBR of AAC residual bitstreams, we adopt single-hop training to simplify the optimization process. The whole training algorithm is summarized in Alg. 3.

\begin{figure*}[htbp]
	\centering  
	\subfigure[PSNR for the reconstructed images.]{
		\includegraphics[width=0.32\linewidth]{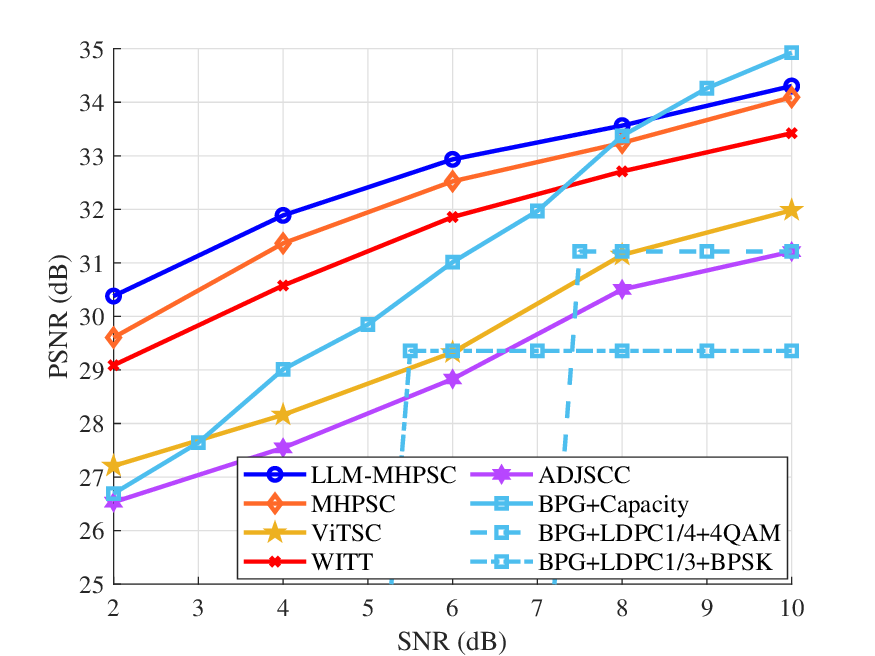}}
	\subfigure[MS-SSIM for the reconstructed images.]{
		\includegraphics[width=0.32\linewidth]{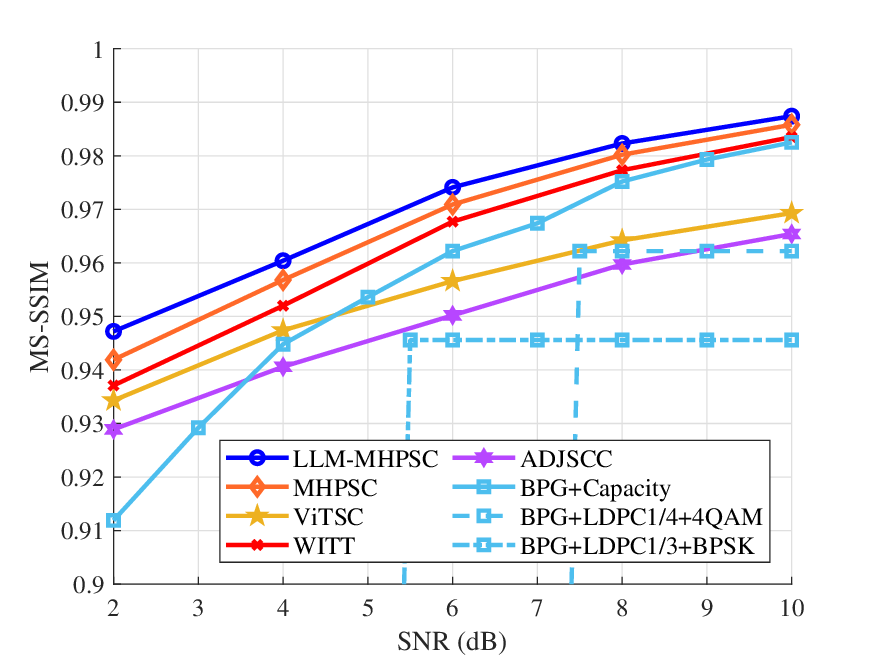}}
	\subfigure[LPIPS for the reconstructed images.]{
		\includegraphics[width=0.32\linewidth]{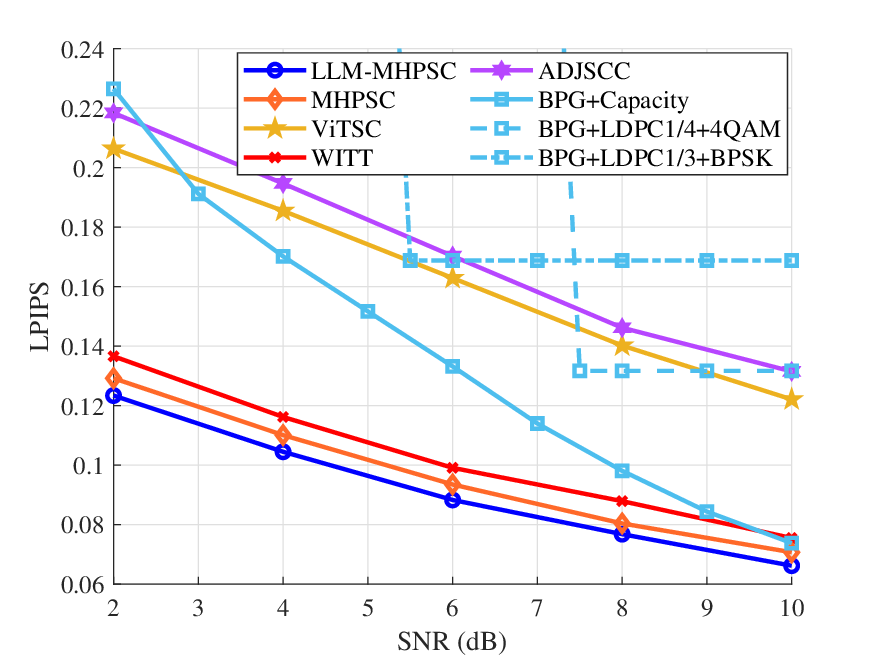}}
	\caption{Quality of the reconstructed images versus the SNRs under Rayleigh fading channels (CBR = 0.035, N = 10).}
	\label{fig_6}
\end{figure*}
\begin{figure*}[htbp]
	\centering  
	\subfigure[PSNR for the reconstructed images.]{
		\includegraphics[width=0.32\linewidth]{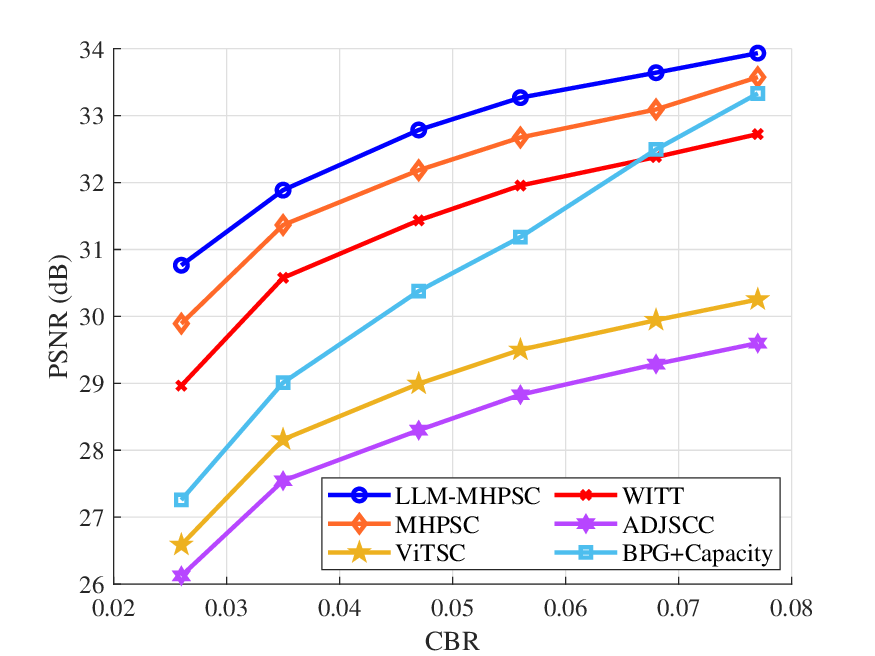}}
	\subfigure[MS-SSIM for the reconstructed images.]{
		\includegraphics[width=0.32\linewidth]{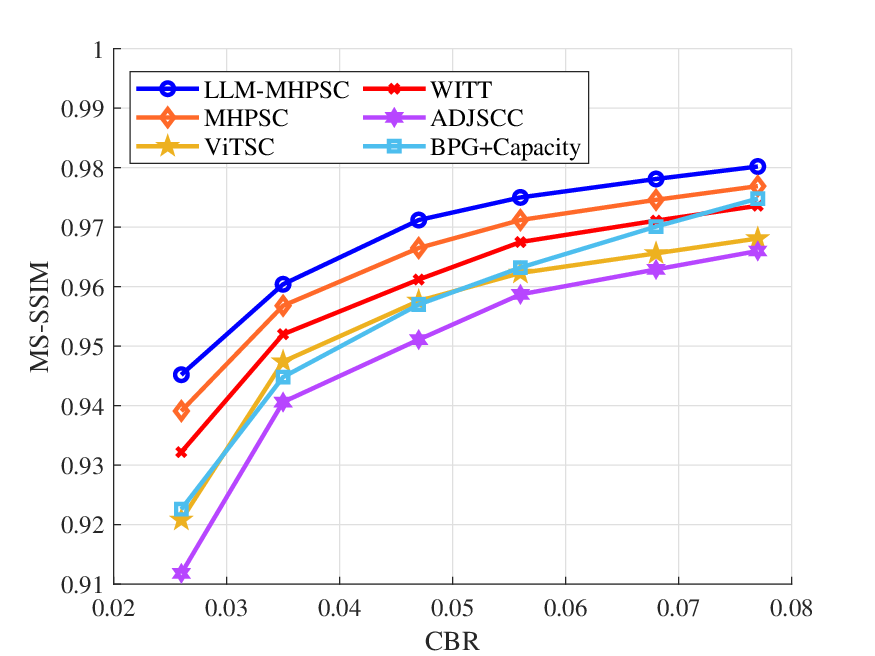}}
	\subfigure[LPIPS for the reconstructed images.]{
		\includegraphics[width=0.32\linewidth]{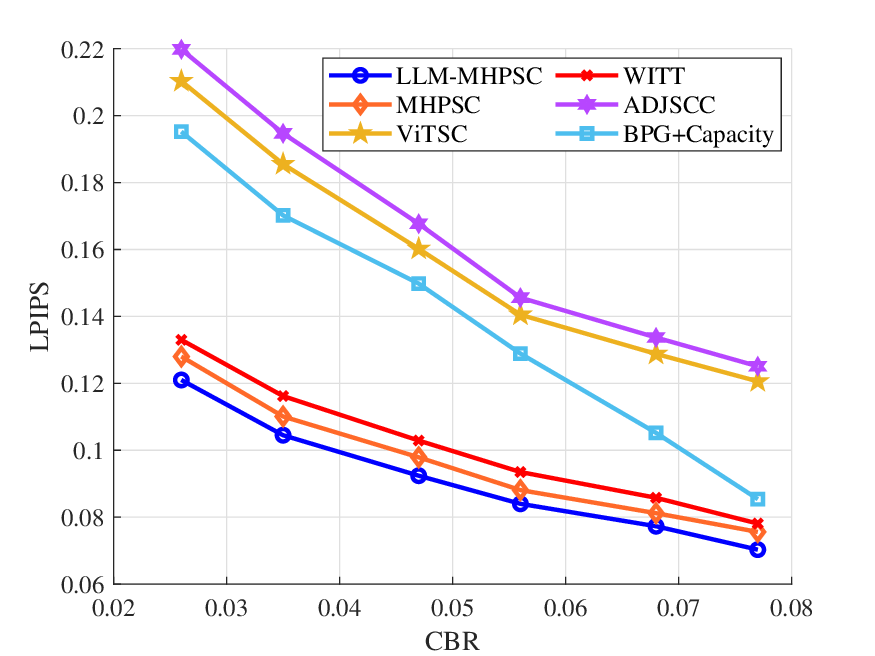}}
	\caption{Quality of the reconstructed images versus the CBRs under Rayleigh fading channels (SNR = 4 dB, N = 10).}
	\label{fig_7}
\end{figure*}

\section{Numerical Results}
In this section, we present numerical results to evaluate the effectiveness of proposed LLM-MHPSC.

\subsection{Experimental Setups}
\subsubsection{Datasets}

For the wireless image semantic transmission, we quantify the performances of proposed LLM-MHPSC versus other benchmarks over the UDIS-D \cite{UDIS-D} dataset, which contains over 10000 real-world images.

\subsubsection{Model Deployment Details}
We employ WITT \cite{WITT} for the semantic link transmission of LLM-MHPSC, where WITT \cite{WITT} is a typical JSCC-based wireless image semantic communication framework. We use variable learning rates, which decrease step-by-step from 1e-4 to 2e-5 for main backbone training while fixed learning rate 1e-4 for LLM-RTO finetuning. The batchsize is set as 16. $\boldsymbol{\lambda}_{\mathrm{init}}=(0.2,0.2,0.1)$. $P'$ is set as 0.5. $C_{\mathrm{th}}$ is set as 0.5. $\gamma$ is set as 1.15. $\alpha$ is set as 0.1. The hop number during training is 4. The whole framework is optimized with Adam \cite{Adam} algorithm. Experiments are conducted in RTX5090 GPUs with Pytorch2.6.0.

\subsubsection{Comparison Benchmarks}
In the experiments, several benchmarks are given as below

$\textbf{MHPSC}$: The multi-hop parallel image semantic communication framework \cite{mhpsc} with CNN-based residual estimator.

$\textbf{ViTSC}$: The DL-empowered semantic communication framework \cite{mhdeepsc} in multi-hop wireless image transmission scenarios.

$\textbf{WITT}$: The wireless image transmission transformer \cite{WITT}.

$\textbf{ADJSCC}$: The adaptive deep JSCC scheme in \cite{ADJSCC} with the CNN structure, directly blending SNR as side information with original features in attention modules.

$\textbf{BPG+Capacity}$: The traditional coding transmission scheme with the Better Portable Graphics (BPG) \cite{bpg} coder as the source coding and the channel capacity achieving code as the channel coding scheme.

$\textbf{BPG+LDPC+QAM/BPSK}$: The traditional transmission scheme with BPG and 5G LDPC \cite{sionna}, enhanced by the MMSE equalization and interleave for combating the Rayleigh fading channels. QAM or binary phase shift keying (BPSK) is utilized as the modulation scheme.

Note that all the DL-based schemes employ the recursive training method. BPG coder is utilized through \cite{bpg} while 5G LDPC and QAM are deployed aided by sionna \cite{sionna}. For the LLM-MHPSC and MHPSC, the capacity achieving code is adopted for the compensated residual link.

\subsubsection{Evaluation Metrics}

We leverage the widely used pixel-wise metric peak signal-to-noise ratio (PSNR), perceptual-level multi-scale structural similarity (MS-SSIM) and learned perceptual image patch similarity (LPIPS) as measurements for the reconstructed image quality.  

Since we consider the multi-hop scenario, the average CBR budget among hops is considered for evaluation. Here CBR refers to the ratio between the transmitted symbol length and the original symbol length as defined in \cite{djscc}. As such, the average CBR among $N$ hops is formulated as
\begin{align}
	CBR=\frac{1}{N}\sum_{n=1}^{N}\frac{L+L^r_n}{H\times W\times 3},
\end{align}
where the codeword length $L^r_n$ of compensation link is varied according to channel conditions and the hop selection choice.

\begin{table*}[htbp]
	\centering
	\caption{Ablations of proposed modules with full LLM-MHPSC as anchor
		($N=20$, average SNR = 6.1 dB, $\frac{L}{H\times W\times 3}=0.0208$).}
	\label{table1}
	\renewcommand{\arraystretch}{1.15}
	\setlength{\tabcolsep}{1.6pt}
	\footnotesize
	
	\begin{tabular}{c c c c c c c c c c c c}
		\toprule
		\multirow{2}{*}{Ablation Terms}
		& \multicolumn{3}{c}{LLM Input Embeddings}
		& \multicolumn{1}{c}{Rate Adjustment}
		& \multicolumn{2}{c}{Hop Selection}
		& \multicolumn{1}{c}{RTO Backbone}
		& \multicolumn{3}{c}{LLM Internal Mechanism}
		& \multirow{2}{*}{\textbf{LLM-MHPSC}} \\
		\cmidrule(lr){2-4}
		\cmidrule(lr){5-5}
		\cmidrule(lr){6-7}
		\cmidrule(lr){8-8}
		\cmidrule(lr){9-11}
		& w/o $V_1(\cdot)$
		& w/o $V_2(\cdot)$
		& w/o $S_a(\cdot,\cdot,\cdot)$
		& w/o RA
		& w/o HS
		& w/ FHS
		& CNN
		& w/o FT
		& w/ TI
		& Masked SA
		& \\
		\midrule
		
		PSNR (dB)
		& \makecell{30.98\\{\color{blue}(-0.16$\%$)}}
		& \makecell{30.95\\{\color{blue}(-0.26$\%$)}}
		& \makecell{31.01\\{\color{blue}(-0.06$\%$)}}
		& \makecell{31.09\\{\color{red}(+0.19$\%$)}}
		& \makecell{31.41\\{\color{red}(+1.22$\%$)}}
		& \makecell{31.11\\{\color{red}(+0.26$\%$)}}
		& \makecell{31.06\\{\color{red}(+0.10$\%$)}}
		& \makecell{30.98\\{\color{blue}(-0.16$\%$)}}
		& 31.03
		& 31.03
		& \textbf{31.03} \\
		
		Extra CBR
		& \makecell{0.0044\\{\color{red}(+7.3$\%$)}}
		& \makecell{0.0046\\{\color{red}(+12.2$\%$)}}
		& \makecell{0.0049\\{\color{red}(+19.5$\%$)}}
		& \makecell{0.0048\\{\color{red}(+17.1$\%$)}}
		& \makecell{0.0060\\{\color{red}(+46.3$\%$)}}
		& \makecell{0.0051\\{\color{red}(+24.4$\%$)}}
		& \makecell{0.0093\\{\color{red}(+126.8$\%$)}}
		& \makecell{0.0047\\{\color{red}(+14.6$\%$)}}
		& \makecell{0.0052\\{\color{red}(+26.8$\%$)}}
		& \makecell{0.0053\\{\color{red}(+29.2$\%$)}}
		& \textbf{0.0041} \\
		
		Runtime (s)
		& \makecell{0.197\\{\color{blue}(-2.5$\%$)}}
		& \makecell{0.201\\{\color{blue}(-0.5$\%$)}}
		& \makecell{0.201\\{\color{blue}(-0.5$\%$)}}
		& 0.202
		& \makecell{0.616\\{\color{red}(+204.9$\%$)}}
		& \makecell{0.256\\{\color{red}(+26.7$\%$)}}
		& \makecell{0.166\\{\color{blue}(-17.8$\%$)}}
		& 0.202
		& \makecell{0.263\\{\color{red}(+30.2$\%$)}}
		& \makecell{0.207\\{\color{red}(+2.5$\%$)}}
		& \textbf{0.202} \\
		
		\bottomrule
	\end{tabular}
\end{table*}

\subsection{Results Analysis}

\subsubsection{Performance for Different SNRs}
We first evaluate the transmission robustness of LLM-MHPSC under Rayleigh fading channels with a specific CBR. As shown in Fig. \ref{fig_6}(a), it can be observed that LLM-MHPSC outperforms all other benchmarks. Compared to other DL-based schemes, LLM-MHPSC outperforms WITT for over 1 dB in terms of PSNR on average, where the gap slightly increases as the SNR value decreases. The performance gains over ViTSC and ADJSCC are also evident, especially in the low-SNR regime. This trend verifies that dual link-based LLM-MHPSC is more robust to the channel interference than single link-based schemes. Compared to MHPSC, which is also a dual-link transmission scheme, the gap between LLM-MHPSC and MHPSC becomes large as SNR decreases. As the performance gain mainly derives from the stronger compression performance of LLM-RTO, LLM-MHPSC can save more residual-link bandwidth than MHPSC in the low SNR regime. For traditional separated coding schemes, BPG with channel capacity achieving code is adopted for providing upper bound for SSCC scheme. LLM-MHPSC significantly outperforms `BPG+Capacity' in low regimes, e.g., 2-6 dB. When the SNR becomes large, LLM-MHPSC also provides comparable performance against `BPG+Capacity'. For `BPG+LDPC' scheme, 1/4 code rate LDPC with 4QAM and 1/3 code rate LDPC with BPSK are set, which reflect corresponding anti-noise performances under different SNR levels. Compared to traditional schemes, LLM-MHPSC provides much more performance gain and stability since traditional schemes suffer from severe cliff effects. As shown in Fig. \ref{fig_6}(b) and Fig. \ref{fig_6}(c), the DL-based schemes achieve better reconstruction results in terms of MS-SSIM and LPIPS compared to traditional SSCC schemes, which means that the satisfying visual perception quality is ensured through extracting semantics inside raw images. Compared to other schemes, LLM-MHPSC preserves even much more high frequency image details. These results demonstrate the effectiveness of LLM-MHPSC for multi-hop transmission under Rayleigh fading channels with various noise intensities.

\subsubsection{Performance for Different CBRs}
Then we evaluate the bandwidth efficiency of LLM-MHPSC with SNR = 4 dB. As shown in Fig. \ref{fig_7}(a), LLM-MHPSC achieves significant performance gain compared to other schemes. Compared to WITT, the performance gap increases as the CBR decreases. This gain mainly stems from the intrinsic distortion accumulation problem in multi-hop scenarios. Transmitting semantics with small CBR results in low quality for image reconstruction. However, residual compensation mitigates the distortion at the end of each hop, greatly alleviating the whole distortion for multiple hops. For perceptual quality metrics in Fig. \ref{fig_7}(b) and Fig. \ref{fig_7}(c), LLM-MHPSC retains relatively satisfying visual quality for a wide range of CBRs compared to other schemes. 

\begin{figure}[htbp]
	\centering
	\includegraphics[width=3.4in]{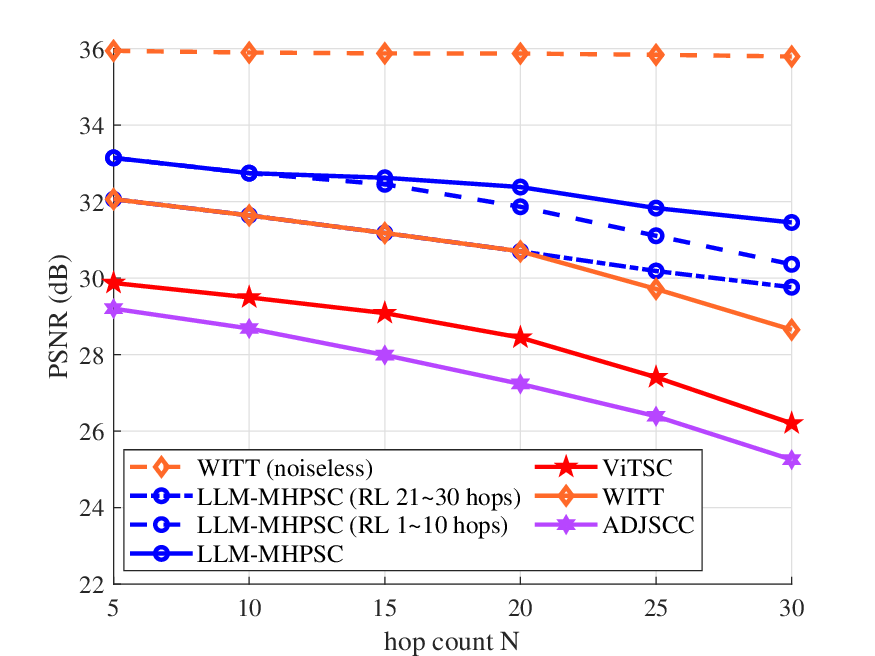}
	\caption{Performance of different hop numbers. (SNR = 8 dB)}
	\label{fig_8}
\end{figure}

\begin{table*}[htbp]
	\centering
	\caption{Different compensation hop counts versus hop selection with three typical channel variation modes. Parentheses indicate PSNR gains over $N_c=0$ and Extra-CBR reductions relative to $N_c=20$.}
	\label{table2}
	\renewcommand{\arraystretch}{1.15}
	\setlength{\tabcolsep}{1.6pt}
	\footnotesize
	
	\begin{tabular}{c c c c c c c c c c c c c c}
		\toprule
		
		\multicolumn{1}{c}{Channel Mode} 
		& \multicolumn{12}{c}{Random Channel Quality Variation ($N=20$, average SNR = 6.1 dB, $\frac{L}{H\times W\times 3}=0.0208$)} \\
		\cmidrule(lr){1-1}\cmidrule(lr){2-14}
		
		$N_c$ 
		& 0 & 2 & \textbf{2.41 (AHS)} & 4 & 4.81 (FHS) & 6 & 8 & 10 & 12 & 14 & 16 & 18 & 20 \\
		PSNR (dB)
		& 29.17
		& \makecell{29.57\\{\color{red}(+0.40)}}
		& \makecell{\textbf{31.03}\\{\color{red}(\textbf{+1.86})}}
		& \makecell{29.79\\{\color{red}(+0.62)}}
		& \makecell{31.11\\{\color{red}(+1.94)}}
		& \makecell{29.96\\{\color{red}(+0.79)}}
		& \makecell{30.01\\{\color{red}(+0.84)}}
		& \makecell{30.33\\{\color{red}(+1.16)}}
		& \makecell{30.68\\{\color{red}(+1.51)}}
		& \makecell{30.98\\{\color{red}(+1.81)}}
		& \makecell{31.17\\{\color{red}(+2.00)}}
		& \makecell{31.28\\{\color{red}(+2.11)}}
		& \makecell{31.41\\{\color{red}(+2.24)}}\\
		Extra CBR
		& \makecell{0\\{\color{blue}(-100.0$\%$)}}  
		& \makecell{0.0004\\{\color{blue}(-93.3$\%$)}} 
		& \makecell{\textbf{0.0041}\\{\color{blue}(\textbf{-31.7$\%$})}} 
		& \makecell{0.0010\\{\color{blue}(-83.3$\%$)}} 
		& \makecell{0.0051\\{\color{blue}(-15.0$\%$)}} 
		& \makecell{0.0027\\{\color{blue}(-55.5$\%$)}} 
		& \makecell{0.0030\\{\color{blue}(-50.0$\%$)}} 
		& \makecell{0.0034\\{\color{blue}(-43.3$\%$)}} 
		& \makecell{0.0041\\{\color{blue}(-31.7$\%$)}} 
		& \makecell{0.0046\\{\color{blue}(-23.3$\%$)}} 
		& \makecell{0.0052\\{\color{blue}(-13.3$\%$)}} 
		& \makecell{0.0054\\{\color{blue}(-10.0$\%$)}} 
		& 0.0060
		\\
		\midrule
		
		\multicolumn{1}{c}{Channel Mode}
		& \multicolumn{12}{c}{Abrupt Channel Variation ($N=20$, average SNR = 6.8 dB, $\frac{L}{H\times W\times 3}=0.0208$)} \\
		\cmidrule(lr){1-1}\cmidrule(lr){2-14}
		
		$N_c$ 
		& 0 & 2 & \textbf{2.34 (AHS)} & 4 & 4.43 (FHS) & 6 & 8 & 10 & 12 & 14 & 16 & 18 & 20 \\
		PSNR (dB)
		& 30.21
		& \makecell{30.45\\{\color{red}(+0.24)}}
		& \makecell{\textbf{31.68}\\{\color{red}(\textbf{+1.47})}}
		& \makecell{30.63\\{\color{red}(+0.42)}}
		& \makecell{31.81\\{\color{red}(+1.60)}}
		& \makecell{30.69\\{\color{red}(+0.48)}}
		& \makecell{30.88\\{\color{red}(+0.67)}}
		& \makecell{31.08\\{\color{red}(+0.87)}}
		& \makecell{31.24\\{\color{red}(+1.03)}}
		& \makecell{31.34\\{\color{red}(+1.13)}}
		& \makecell{31.64\\{\color{red}(+1.43)}}
		& \makecell{31.78\\{\color{red}(+1.57)}}
		& \makecell{32.10\\{\color{red}(+1.89)}}\\
		Extra CBR
		& \makecell{0\\{\color{blue}(-100.0$\%$)}}
		& \makecell{0.0006\\{\color{blue}(-88.0$\%$)}} 
		& \makecell{\textbf{0.0031}\\{\color{blue}(\textbf{-38.0$\%$)}}} 
		& \makecell{0.0012\\{\color{blue}(-76.0$\%$)}} 
		& \makecell{0.0040\\{\color{blue}(-20.0$\%$)}} 
		& \makecell{0.0021\\{\color{blue}(-58.0$\%$)}} 
		& \makecell{0.0025\\{\color{blue}(-50.0$\%$)}} 
		& \makecell{0.0030\\{\color{blue}(-40.0$\%$)}} 
		& \makecell{0.0033\\{\color{blue}(-34.0$\%$)}} 
		& \makecell{0.0035\\{\color{blue}(-30.0$\%$)}} 
		& \makecell{0.0042\\{\color{blue}(-16.0$\%$)}} 
		& \makecell{0.0045\\{\color{blue}(-10.0$\%$)}} 
		& 0.0050\\
		\midrule
		
		\multicolumn{1}{c}{Channel Mode}
		& \multicolumn{12}{c}{Stable Channel Quality ($N=20$, average SNR = 8.3 dB, $\frac{L}{H\times W\times 3}=0.0208$)} \\
		\cmidrule(lr){1-1}\cmidrule(lr){2-14}
		
		$N_c$ 
		& 0 & 2 & \textbf{2.20 (AHS)} & 3.67 (FHS) & 4 & 6 & 8 & 10 & 12 & 14 & 16 & 18 & 20 \\
		PSNR (dB)
		& 30.84
		& \makecell{31.01\\{\color{red}(+0.17)}}
		& \makecell{\textbf{31.87}\\{\color{red}(\textbf{+1.03})}}
		& \makecell{32.01\\{\color{red}(+1.17)}}
		& \makecell{31.08\\{\color{red}(+0.24)}}
		& \makecell{31.28\\{\color{red}(+0.44)}}
		& \makecell{31.30\\{\color{red}(+0.46)}}
		& \makecell{31.45\\{\color{red}(+0.61)}}
		& \makecell{31.60\\{\color{red}(+0.76)}}
		& \makecell{31.77\\{\color{red}(+0.93)}}
		& \makecell{31.89\\{\color{red}(+1.05)}}
		& \makecell{32.02\\{\color{red}(+1.18)}}
		& \makecell{32.35\\{\color{red}(+1.51)}}\\
		Extra CBR
		& \makecell{0\\{\color{blue}(-100.0$\%$)}}
		& \makecell{0.0003\\{\color{blue}(-91.4$\%$)}}
		& \makecell{\textbf{0.0021}\\{\color{blue}(\textbf{-40.0}$\%$)}}
		& \makecell{0.0026\\{\color{blue}(-25.7$\%$)}}
		& \makecell{0.0005\\{\color{blue}(-85.7$\%$)}}
		& \makecell{0.0013\\{\color{blue}(-62.9$\%$)}}
		& \makecell{0.0016\\{\color{blue}(-54.3$\%$)}}
		& \makecell{0.0019\\{\color{blue}(-45.7$\%$)}}
		& \makecell{0.0022\\{\color{blue}(-37.1$\%$)}}
		& \makecell{0.0025\\{\color{blue}(-28.6$\%$)}}
		& \makecell{0.0027\\{\color{blue}(-22.9$\%$)}}
		& \makecell{0.0030\\{\color{blue}(-14.3$\%$)}}
		& 0.0035\\
		
		\bottomrule
	\end{tabular}
\end{table*}

\subsubsection{Ablation Study}
The ablation results are reported in Tab.~\ref{table1}. The experiments are conducted under a 20-hop setting, where the SNR varies from 2 dB to 10 dB with an average value of 6.1 dB, and the semantic link CBR is fixed as 0.0208. Extra CBR is the residual link CBR cost. Runtime denotes the average inference time for transmitting one image across $N$ hops. We first evaluate the contribution of LLM input embeddings. Removing $V_1(\cdot)$ or $V_2(\cdot)$ degrades PSNR, indicating that both reference and residual details are useful for residual distribution estimation to some extent. Without $S_a(\cdot,\cdot,\cdot)$, PSNR slightly decreases while CBR increases, verifying the importance of CSI and hop-aware information. For rate adjustment and hop selection, removing RA slightly improves PSNR but increases CBR, showing that adaptive masking effectively reduces residual link overhead with marginal losses. Without HS, higher PSNR is achieved at the cost of much larger CBR and runtime, since the compensation link is activated more frequently, likewise the fixed hop selection (FHS) strategy. These results confirm that the gate-based adaptive hop selection achieves a better balance among reconstruction quality, bandwidth overhead, and computational cost. We then compare LLM-RTO with an alternative RTO backbone. Replacing LLM-RTO with the CNN-based residual estimator in~\cite{mhpsc} slightly improves PSNR but substantially increases CBR, indicating that CNN-based residual modeling relies on more transmitted residual bits to achieve high reconstruction quality. In contrast, LLM-RTO provides more compression-efficient residual distribution estimation. Finally, the LLM internal mechanism ablation verifies the roles of fine-tuning, token interaction, and self-attention. Without FT, both PSNR and CBR become worse, showing that task-specific adaptation is necessary. With token isolation (TI), where reference, residual, and SEI embeddings are processed without cross-token interaction, PSNR remains similar but CBR and runtime increase, indicating that heterogeneous token interaction improves compression efficiency. Masking full self-attention also increases CBR, confirming the benefit of global contextual modeling. Overall, these results show that the gain of LLM-RTO comes from the joint effect of partial fine-tuning, heterogeneous token interaction, and full self-attention, rather than merely from using a large Transformer backbone. To conclude, LLM-MHPSC achieves a favorable PSNR-CBR-Runtime trade-off due to the cooperative design of both LLM-based structure and hop selection strategy.

\subsubsection{Performances of different numbers of hops}
After that, we evaluate performances of different hop counts. In Fig. \ref{fig_8}, $N$ ranges from 5 to 30, covering low frequency and high frequency multi-hop conditions. It is clearly to observe that LLM-MHPSC surpasses all other DL-based schemes in terms of PSNR for various hop counts. For proposed LLM-MHPSC, less performance degradation is presented with much larger hop number such as $N=30$. While for other schemes, although the same recursive training method is adopted to promote each hop aware of previous hops, the reconstructed image quality degrades seriously with more hops. We further evaluate LLM-MHPSC by deploying its compensation link over different hop sequences. Specifically, RL 1$\sim$10 refers to the application of residual link only to the initial 10 hops, while RL 21$\sim$30 for only the final 10 hops. The results indicate that compensating the initial hops maintains stable performance for approximately 15 hops, after which it degrades as the hop count increases. In contrast, compensating the final hops gradually mitigates distortion accumulation, with performance improving as the hop count rises. However, it does not achieve performances as good as compensation for the early 10 hops. This trend demonstrates that early hop compensation is commonly more important than the late hop compensation. Since LLM-MHPSC adapts hop selection strategy, it can achieve better performance with less compensation links. Finally, WITT (noiseless) refers to the noiseless semantic link, which means that only semantic loss exists. It means that proposed LLM-MHPSC mainly mitigates accumulated distortion from the wireless channel of each link. With the above analysis, LLM-MHPSC is able to perform robust wireless image transmission in multi-hop scenarios.

\begin{figure*}[htbp]
	\centering
	\includegraphics[width=4.4in]{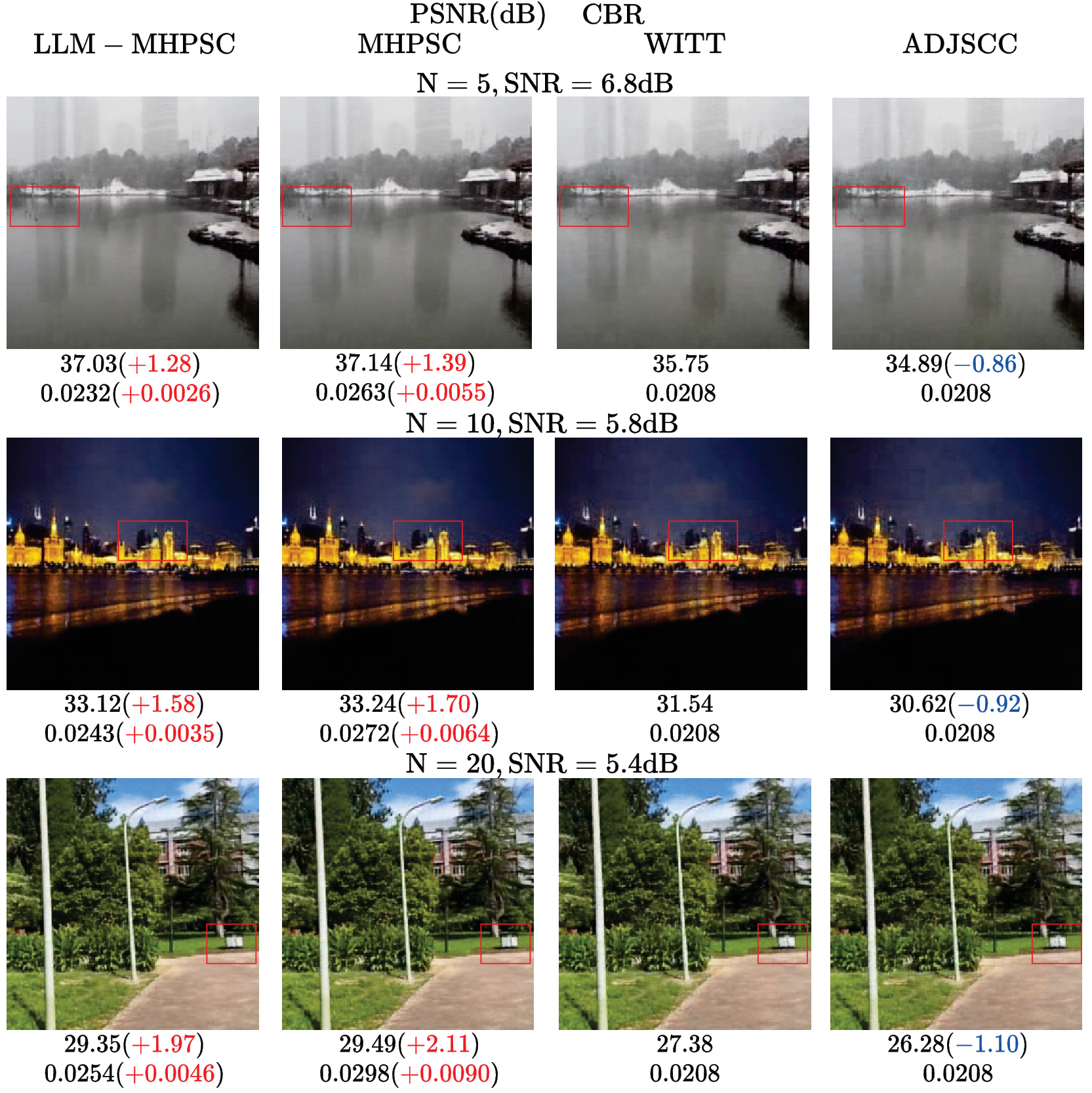}
	\caption{Visual results of LLM-MHPSC and other benchmarks. The baseline is WITT, which is the semantic link.}
	\label{fig_9}
\end{figure*}

\subsubsection{Performances of Hop Selection under Varying Channel Conditions}

We further evaluate LLM-MHPSC under different compensated hop counts and channel variation modes, as reported in Tab. \ref{table2}. Three representative cases are considered: random channel quality variation, abrupt channel variation with dramatic inter-hop degradation, and stable channel quality where the SNR varies from 6 dB to 10 dB. Here, $N_c$ denotes the number of activated compensation hops in the sequential compensation scheme, where only the first $N_c$ hops are compensated except for AHS and FHS. In contrast, AHS denotes the proposed adaptive hop selection strategy, which determines compensation activation according to channel conditions, historical compensation states, and the compensation budget. FHS denotes the fixed hop selection strategy.

From Tab. \ref{table2}, increasing $N_c$ generally improves PSNR but also increases CBR, showing the trade-off between reconstruction quality and residual-link overhead. Sequentially compensating more early hops can effectively suppress distortion propagation, but the gain becomes less efficient when excessive hops are compensated. This indicates that activating the residual link at every hop is unnecessary for achieving a favorable CBR-PSNR balance. The proposed AHS achieves strong performance with only a small number of activated hops. Under random channel variation, AHS activates 2.41 hops on average and achieves 31.04 dB, which is much higher than sequential compensation with $N_c=4$ and close to FHS with fewer activations and lower CBR. Under abrupt channel variation, AHS achieves 31.68 dB with only 2.34 activated hops, demonstrating its ability to identify critical hops suffering from severe channel degradation. Under stable channel quality, AHS also provides a clear PSNR gain over sequential compensation with similar or fewer activated hops. These results verify that the proposed AHS can allocate compensation more efficiently than simple sequential activation. Overall, AHS provides an effective quality-overhead trade-off across different channel variation modes. By selectively activating the residual link only at critical hops, LLM-MHPSC reduces unnecessary compensation while preserving strong robustness against multi-hop distortion accumulation.

\subsubsection{Visualized Results}
We present visualization results for multi-hop transmission in Fig. \ref{fig_9}, where the PSNR performance and corresponding CBR are illustrated. WITT is adopted as the baseline, and comparisons under different hop counts $N$ and average SNR levels are provided. In general, the proposed LLM-MHPSC outperforms both WITT and ADJSCC in terms of image reconstruction quality, with only a moderate increase in bandwidth cost. For instance, at $N$=5 and  SNR=6.8 dB, LLM-MHPSC introduces merely 11$\%$ additional bandwidth for the residual compensation link, while achieving a PSNR gain of 1.28 dB. In contrast, although MHPSC yields a slightly higher PSNR gain, its extra bandwidth cost surges to 26$\%$, leading to a less favorable CBR-PSNR tradeoff compared with LLM-MHPSC. This phenomenon arises because MHPSC employs full compensation across every hop, resulting in excessive overhead. Furthermore, the performance gap between LLM-MHPSC and WITT (semantic link) widens as the hop count $N$ increases, which validates the effectiveness of the proposed plug-and-play compensation link in mitigating multi-hop distortion accumulation.

\subsubsection{Complexity analysis}
\begin{table}[htbp]
	\centering
	\caption{Evaluation of complexity and computation cost.}
	\label{table3}
	
	\begin{tabular}{|c|c|c|c|}  
		\hline 
		& & &\\[-6pt] 
		Metric&FLOPs (G)&Runtime (s)&Parameters (M) \\
		\hline
		& & &\\[-6pt]  
		\textbf{LLM-MHPSC}&\textbf{14.01}&\textbf{0.103}&\textbf{1741.86} \\
		\hline
		& & &\\[-6pt]  
		LLM-MHPSC (FHS)&23.29&0.124&1741.86 \\
		\hline
		& & &\\[-6pt]  
		LLM-MHPSC (w/o HS)&70.28&0.251&1741.86 \\
		\hline
		& & &\\[-6pt]  
		MHPSC&10.03&0.141&20.24 \\
		\hline
		& & &\\[-6pt] 
		WITT&6.62&0.074&13.99\\
		\hline
	\end{tabular}
\end{table}

Finally, to evaluate the feasibility of LLM-MHPSC, we analyze the complexity and computation cost. FLOPs represent the number of floating-point operations required for inference. Parameters is the total parameters of a model. As shown in Tab. \ref{table3}, due to the LLM deployment, the required computation complexity is much higher than the MHPSC and WITT. However, when it comes to the actual runtime, the gap narrows greatly for the utilization of lightweight Qwen-1.7B and hop selection strategy. Compared to MHPSC, LLM-MHPSC performs a lower runtime and achieves better compensation performances. When hop selection strategy is not adopted, runtime tends to be significantly longer due to the employment of LLM-RTO for every hop. WITT has the minimum runtime compared to LLM-MHPSC and MHPSC. However, it lacks robustness for accumulated distortion among multiple hops, especially for the low SNR or high hop count regimes. This reflects a trade-off between runtime cost and multi-hop transmission performance. However, LLM-MHPSC provides a plug-and-play solution without affecting the semantic link. We believe that with more lightweight LLMs, e.g. 0.6B, along with up-to-date LLM acceleration scheme, the runtime can be further reduced.

\section{Conclusion}
This paper proposed LLM-MHPSC, a novel framework addressing distortion accumulation in multi-hop wireless image transmission. LLM-MHPSC mitigates information loss across hops while minimizing bandwidth overhead, by introducing a residual compensation link with coarse-to-fine compression. The key innovation, LLM-RTO, leverages fine-tuned large language models to jointly optimize residual distribution estimation and enable CSI- and hop-aware rate adjustment. Together with an adaptive hop selection strategy that balances performance and computational cost, the proposed framework allows plug-and-play integration with existing single-link semantic communication schemes. Extensive experiments demonstrate that LLM-MHPSC consistently outperforms state-of-the-art semantic and traditional coding schemes across various SNR levels, bandwidth ratios, and hop counts, achieving robust image transmission with only marginal overhead increase. While this work focuses on mitigating distortion accumulation through residual compensation, an interesting direction for future research is to further enhance the communication efficiency of the residual compensation link itself, for example by enabling joint source-channel coding and superposition coding with the semantic communication links.

\end{document}